\UseRawInputEncoding
\documentclass[reprint,prc,amsmath,amssymb,superscriptaddress,noffotinbib,preprintnumbers]{revtex4-1}
\usepackage{url,color,colordvi,epsfig,pstricks}
\usepackage[colorlinks=true, citecolor=blue, filecolor=blue, linkcolor=blue, urlcolor=blue, pdftex]{hyperref}
\usepackage{hyperref}
\usepackage{cleveref}
\usepackage{graphicx,bm,dcolumn}
\usepackage{natbib}

\hypersetup{
    colorlinks=true, 
    linktoc=all, 
    linkcolor=blue, 
    citecolor=blue}
    
\begin{document}

\preprint{FERMILAB-PUB-25-0748-PPD, WSU-HEP-2502}

\title{Prospects for Exploring Non-Standard Neutrino Properties with Argon-Based CEvNS Experiments}

\author{S.~Carey}\email{samcarey@wayne.edu}
\affiliation{Department of Physics and Astronomy, Wayne State University, Detroit, Michigan 48201, USA}

\author{V.~Pandey}\email{vpandey@fnal.gov}
\affiliation{Fermi National Accelerator Laboratory, Batavia, Illinois 60510, USA}

\begin{abstract}
Coherent elastic neutrino--nucleus scattering (CEvNS) provides a powerful framework for testing the Standard Model (SM) and searching for new physics at low energies. In this work, we examine the prospects for argon-based CEvNS experiments at stopped-pion sources to perform precision measurements of weak interactions and probe non-standard neutrino properties. Our study focused on the CENNS-10 and CENNS-750 detectors at the Spallation Neutron Source at Oak Ridge National Laboratory, the Coherent Captain Mills (CCM) detector at Los Alamos National Laboratory, and the proposed PIP2-BD detector at Fermilab's Facility for Dark Matter Discovery (F2D2). Using realistic neutrino fluxes and detector configurations corresponding to these facilities, we evaluate event rates and sensitivities to a range of observables. Within the SM, argon-based CEvNS detectors enable precision tests of electroweak parameters, including the weak mixing angle, at momentum transfers well below the electroweak scale. We also investigate the sensitivity of these experiments to neutrino electromagnetic properties, such as the magnetic moment and effective charge radius, as well as to possible non-standard neutrino interactions with quarks. Together, these studies highlight the potential of argon-based CEvNS experiments as a clean and versatile platform for precision exploration of non-standard neutrino properties.
\end{abstract}

\maketitle



\section{Introduction}\label{sec:intro}

The study of neutrino interactions with nuclei has long served as a critical avenue for advancing both particle physics and astrophysics \cite{Formaggio:2012cpf, Balasi:2015dba}. Among these processes, coherent elastic neutrino-nucleus scattering (CEvNS) occupies a special place. First predicted in the mid-1970s as a direct consequence of the Standard Model (SM) \cite{Freedman:1973yd, Drukier:1984vhf}, CEvNS arises when a neutrino scatters coherently off an entire nucleus, leading to an enhanced cross section that scales approximately with the square of the neutron number. Despite its relatively large predicted cross section compared to other low-energy neutrino processes, the detection of CEvNS proved elusive for more than four decades due to the tiny nuclear recoil energies, typically of order tens of keV, that must be measured \cite{Scholberg:2005qs}.

This challenge was overcome in 2017, when the COHERENT collaboration reported the first observation of CEvNS using a CsI detector at the Spallation Neutron Source \cite{COHERENT:2017ipa}. Subsequent measurements with additional detector technologies \cite{CCM:2021leg, Colaresi:2021kus, COHERENT:2021xmm, Ackermann:2025obx, COHERENT:2025vuz}, including liquid argon \cite{COHERENT:2020iec}, have strengthened this discovery and expanded the physics reach of CEvNS. The achievement represented not only a milestone in confirming a long-standing SM prediction, but also the opening of a new experimental frontier for precision neutrino physics at low energies.

The physics potential of CEvNS stems from its unique sensitivity to both SM parameters and possible signatures of new physics. On the SM side, CEvNS enables measurements of the weak mixing angle at low momentum transfer \cite{Papoulias:2017qdn, Canas:2018rng, Cadeddu:2019eta, Miranda:2020tif, Cadeddu:2020lky, Cadeddu:2021ijh, AristizabalSierra:2022axl, AtzoriCorona:2022qrf, AtzoriCorona:2023ktl, AtzoriCorona:2024vhj, Lindner:2024eng}, tests of nuclear form factors, and probes of nuclear structure \cite{Cadeddu:2017etk, Cadeddu:2018izq, Papoulias:2019lfi, AristizabalSierra:2019zmy, Coloma:2020nhf, Rossi:2023brv, VanDessel:2020epd, Pandey:2023arh, Tomalak:2020zfh}. At the same time, CEvNS provides an avenue to explore beyond the Standard Model (BSM) scenarios \cite{Barranco:2005yy, Scholberg:2005qs, Lindner:2016wff, Coloma:2017ncl, Papoulias:2017qdn}. Deviations in recoil spectra or total rates could reveal the presence of non-standard neutrino interactions (NSIs) \cite{Barranco:2005yy, Scholberg:2005qs, Papoulias:2017qdn, Liao:2017uzy, AristizabalSierra:2017joc, Coloma:2017ncl, Coloma:2017egw, Altmannshofer:2018xyo, AristizabalSierra:2018eqm, Proceedings:2019qno, Denton:2022nol, AristizabalSierra:2024nwf}, new mediators \cite{Billard:2013qya, Dent:2016wcr, Lindner:2016wff, Ge:2017mcq, Bauer:2018onh, Abdullah:2018ykz, Billard:2018jnl, AristizabalSierra:2019ykk}, sterile neutrinos \cite{Formaggio:2011jt, Anderson:2012pn, Dutta:2015nlo, Kosmas:2017zbh, Papoulias:2017qdn}, or neutrino electromagnetic properties \cite{Giunti:2014ixa, Kosmas:2015sqa, Kosmas:2015vsa, Cadeddu:2018dux, Miranda:2019wdy, Miranda:2020tif, Giunti:2024gec, AtzoriCorona:2025ibl} such as a magnetic moment or a finite charge radius. In this way, CEvNS serves as both a precision laboratory for weak interactions and a discovery channel for new physics at the MeV scale.

Experimentally, stopped-pion neutrino sources are especially well suited for CEvNS studies \cite{COHERENT:2017ipa, CCM:2021leg, Colaresi:2021kus, COHERENT:2021xmm, Ackermann:2025obx, COHERENT:2025vuz}. These sources produce fluxes of neutrinos in the tens-of-MeV energy range, where the CEvNS cross section is sizable and the recoils remain within the reach of modern low-threshold detectors. The compactness of such detectors allows them to be deployed close to the source, further increasing event rates. As a result, spallation-based facilities have become central to the global CEvNS program. COHERENT at Oak Ridge \cite{COHERENT:2017ipa, COHERENT:2021xmm, COHERENT:2025vuz}, Coherent Captain Mills (CCM) at Los Alamos \cite{CCM:2021leg}, and several proposed efforts \cite{Colaresi:2021kus, Ackermann:2025obx} worldwide are exploiting this opportunity, each with distinct detector technologies and target materials.

Among these, liquid argon (LAr) has emerged as a particularly attractive target. Argon is both abundant and scalable, allowing for the construction of multi-ton detectors, and its nuclear properties provide complementary information to heavier nuclei. Moreover, the development of large LAr detectors in neutrino oscillation experiments, notably the Short-Baseline Neutrino (SBN) program \cite{MicroBooNE:2015bmn} and the Deep Underground Neutrino Experiment (DUNE) \cite{DUNE:2015lol}, has led to significant advances in argon detector technology that can be leveraged for CEvNS applications. Argon-based CEvNS measurements thus offer a unique path to precision tests of weak interactions, while also serving as a proving ground for new physics searches.
Looking ahead, the physics reach of CEvNS will be extended by upcoming facilities, including Fermilab's Facility for Dark Sector Discovery (F2D2)~\cite{Aguilar-Arevalo:2023dai}, which will operate in a beam-dump configuration to produce an intense source of stopped-pion neutrinos and offers the potential to host kiloton-scale experiments. These experiments, in conjunction with global efforts, are poised to deliver unprecedented statistics and precision, enabling detailed studies of SM parameters and strong sensitivity to a broad range of BSM scenarios. In this work, we focus on the prospects of argon-based CEvNS experiments. We present estimates of recoil event rates for current and future detectors, evaluate the sensitivity of these experiments to the weak mixing angle, and explore their reach in probing neutrino electromagnetic properties and NSIs.

The rest of this paper is organized as follows. In Sec. \ref{sec:cross_section}, we outline the formalism for the CEvNS cross section and summarize the different form factors used to characterize the weak structure of the argon nucleus. Sec.~\ref{sec:experiments} outlines the argon-based experiments that form the basis of this analysis. In Sec. \ref{sec:analysis}, we examine constraints on the weak mixing angle, the electromagnetic properties of neutrinos, including the magnetic moment and charge radius, as well as possible non-standard interaction parameters. Finally, the results and our conclusions are summarized in Sec. \ref{sec:conclusions}


\section{CEvNS Cross Section and Form Factors}\label{sec:cross_section}

\begin{table}
\centering
\begin{tabular}{@{}lcc@{}}
\toprule
\textbf{Form Factor} & \textbf{Events/year ($N$)} & \textbf{\(|\Delta N|_{\rm Helm}\)} \\
\hline
Helm   & $5.131\times 10^{6}$ & --- \\
Klein-Nystrand     & $5.117\times 10^{6}$ & $0.28\%$ \\
Van-Dessel \textit{et al.}   & $5.098\times 10^{6}$ & $0.65\%$ \\
Payne \textit{et al.}   & $4.991\times 10^{6}$ & $2.73\%$ \\
Yang \textit{et al.}    & $5.055\times 10^{6}$ & $1.48\%$ \\
Hoferichter \textit{et al.} & $5.089\times 10^{6}$& $0.82\%$\\
\hline
\end{tabular}
\caption{Expected CEvNS events per year at PIP2-BD F2D2 for different nuclear form factor models. The last column shows the percentage difference relative to the Helm form factor prediction.}
\label{tab:events_formfactors}
\end{table}

For an incoming neutrino with energy \(E_\nu\) scattering off an argon nucleus at rest with mass \(M_A\), the differential cross section for CEvNS in the SM is given by
\begin{align}\label{eq:SM_cross_section}
    \left[\frac{d\sigma (E_\nu)}{dT}\right]_{\rm SM} = \frac{G_F^2 Q_W^2 M_A}{4\pi}\left(1-\frac{T}{E_\nu}-\frac{M_A T}{2\,E_\nu^2}\right)F_{\rm Weak}^2 (Q^2).
\end{align}
Where $T$ is the nuclear recoil energy, \(G_F\) is the Fermi coupling constant, and \(Q_W\) is the weak nuclear charge at the tree-level given by, 
\begin{align}\label{eq:weak_charge}
    Q_W = \left[(1-4\,\sin^2\theta_W)Z-N\right],
\end{align}
with \(N,Z\) the number of neutrons and protons, and \(\theta_W\) the weak mixing angle. In the $\overline{\mathrm{MS}}$ scheme, the low-momentum-transfer value is  \(\sin^2\theta_W = 0.23873\) \cite{ParticleDataGroup:2024cfk}. The weak form factor, $F_W(Q^2)$, encapsulates all aspects of the nuclear dynamics in this elastic scattering. 
Several models of the form factor have been proposed in the literature. The Helm \cite{Helm:1956zz} and Klein-Nystrand \cite{Klein:1999} form factors are most commonly used because of their analytic simplicity and reasonable agreement with experimental data. More microscopic nuclear theory approaches, such as Hartree--Fock (HF) \cite{VanDessel:2020epd}, relativistic mean-field (RMF) \cite{Yang:2019pbx}, ab initio next-to-next-to-leading order (NNLO) \cite{Payne:2019wvy}, and chiral Effective Field Theory-Shell Model (EFT-SM)\cite{Hoferichter:2020osn} descriptions are available in the literature. In this work, we adopt the Helm form factor as our baseline, defined as the convolution of uniform spherical density of radius \(R\) with a Gaussian surface of width $s$:
\begin{align}\label{eq:helm}
    F_{\rm Helm}(Q^2) = \frac{3\,j_1(Q\,R)}{Q R}e^{-Q^2\,s^2/2}
\end{align}
where \(j_1\) is the spherical Bessel function of the first kind. The nuclear charge radius \(R^2 = (1.23A^{1/3}-0.6)^2+7/3\pi^2 r_0^2-5s^2\) with \(r_0 = 0.52\) fm and \(s = 0.9\) fm for argon nucleus \cite{Duda:2006uk}.
In Fig.~\ref{fig:cs_cmprsn} of Appendix~\ref{app:ff_comparison}, we show the total CEvNS cross section for $^{40}$Ar as a function of neutrino energy for different nuclear form factor prescriptions. The relative differences between form factor models increase with energy due to the growing sensitivity to nuclear structure effects at higher momentum transfer.

The predicted total event yield per year, assuming 5000 hours of annual operation comparable to SNS running conditions \cite{COHERENT:2022nrm}, is summarized in Table~\ref{tab:events_formfactors}. For each nuclear form factor model, the table also shows the relative deviation with respect to the Helm model, taken as the baseline prediction. These variations introduce a systematic uncertainty of less than 3\%, which is included in the overall systematic uncertainty budget considered in this work.

\begin{table*}
    \centering
    \begin{tabular}{c c c c c c}
    \hline
    \hline
        \textbf{Experiment (Ref.)} & \textbf{$m_T$ (kg)}  & \textbf{\(d\) (m)}  & \textbf{POT (sec$^{-1}$)} & \textbf{Threshold (keVnr)} \\
        \hline
         PIP2-BD at FNAL (\cite{Aguilar-Arevalo:2023dai}) & $100,000$ & $20.0$  & $2.75\times 10^{16}$ & $20$\\
         CENNS10 at ORNL (\cite{COHERENT:2020iec}) & $24$ & $27.5$  & $1.0\times 10^{16}$ & $20$\\
         CENNS750 at ORNL (\cite{COHERENT:2022nrm}) & $610$ & $27.5$ & $7.6\times 10^{15}$ & $20$\\
         CCM at LANL (\cite{CCM:2021leg}) & $10,000$ & $20.0$ & $5.6\times 10^{14}$ & $10$\\
         \hline
    \end{tabular}
    \caption{Key parameters for argon-based CEvNS experiments considered in this work. $m_T$ is the fiducial target mass, $d$ is the distance from the source, and POT is the proton-on-target rate.}
    \label{tab:expt_list}
\end{table*}

\begin{figure*}
	\centering
	\includegraphics[width=0.46\linewidth]{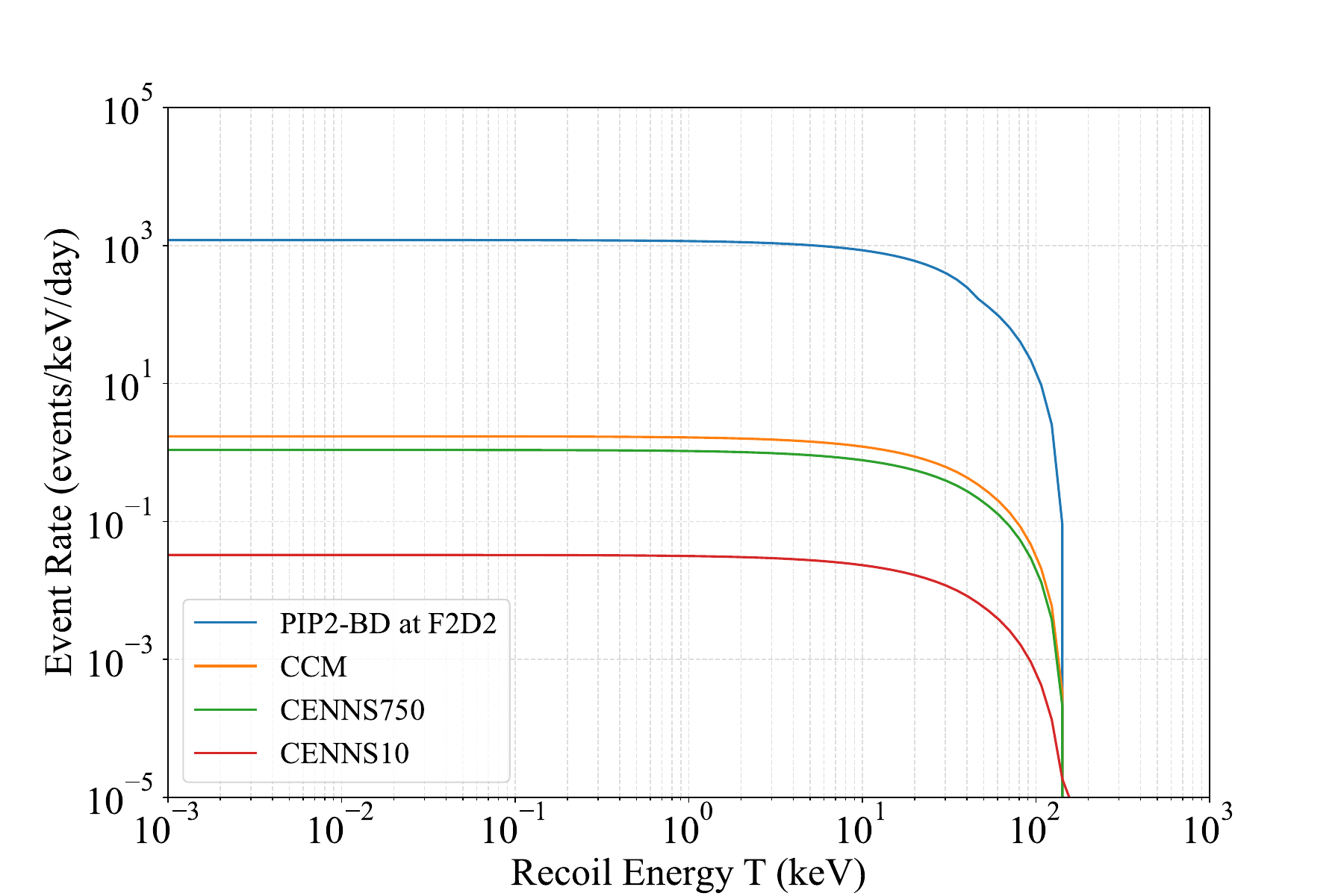}
    \includegraphics[width=0.46\linewidth]{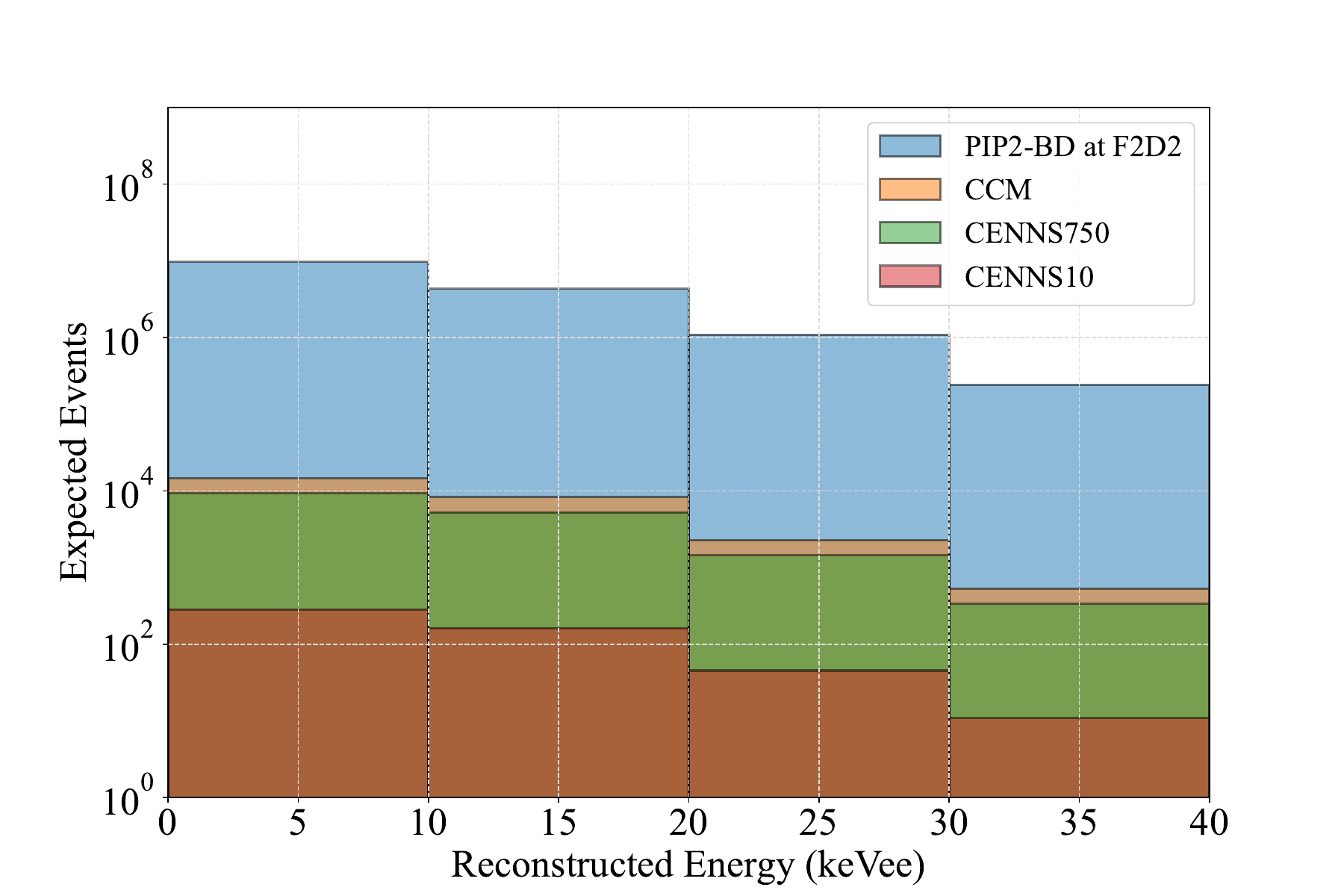}
	\caption{CEvNS event rate per day as a function of the nuclear recoil energy (left) and expected reconstructed event rate accumulated over three years for different argon-based detectors and flux source (right). }
	\label{fig:Rate_Fluxes}
\end{figure*}


\section{Argon-Based CEvNS Experiments}\label{sec:experiments}
With the first observation of CEvNS by the COHERENT collaboration in 2017 \cite{COHERENT:2017ipa}, numerous experimental efforts have since been initiated, or are currently underway, to study CEvNS and possible BSM signatures using stopped--pion as well as reactor neutrino sources. The measurement by COHERENT collaboration, with an exposure of 14.6-308~kg$\cdot$days, observed $134 \pm 22$ nuclear recoil events on CsI using a single photomultiplier tube. This result was consistent with the SM prediction of $178 \pm 43$ events and marked the first confirmation of CEvNS, thereby motivating a wide range of new proposals and experiments employing diverse detector technologies and physics goals. In this work, we focus on liquid argon based CEvNS experiments.
In Table \ref{tab:expt_list}, we list the experiments (and their configurations) involving argon targets that are currently operating \cite{COHERENT:2020iec, CCM:2021leg} and the upgrades proposed for these existing experiments \cite{COHERENT:2022nrm} and a new proposed experiment \cite{Aguilar-Arevalo:2023dai}.   

\textbf{CENNS10 and CENNS750 at ORNL:} The Spallation Neutron Source (SNS) at Oak Ridge National Laboratory (ORNL) provides an intense, pulsed stopped-pion neutrino beam ideal for CEvNS studies \cite{Barbeau:2021exu}. The facility operates with a 1 GeV proton beam at 1.7 MW power \cite{COHERENT:2025vuz} and, after further upgrades, the beam energy and power is expected to reach 1.3 GeV and 2 MW \cite{COHERENT:2021yvp}. The First Target Station (FTS) directs 60 Hz, 400 ns proton pulses onto a liquid mercury target, producing $\pi^+$ that decay at rest ($\pi$-DAR), yielding a prompt $\nu_\mu$ flux followed by delayed $\nu_e$ and $\bar{\nu}_\mu$ neutrinos extending to $\sim$8~$\mu$s due to the 2.2 $\mu$s $\mu^+$ lifetime. This precise timing structure enables suppression of steady-state backgrounds. A second target station (STS) \cite{Asaadi:2022ojm}, now under construction, will increase total beam power to 2.8 MW, further improving sensitivity to rare processes.

The LAr detectors of interest for this work by the COHERENT collaboration at SNS are the CENNS10 and CENNS750 detectors \cite{COHERENT:2022nrm}. The CENNS-10 detector (28 kg LAr, with an active mass of ∼24 kg) observed CEvNS in 2021 \cite{COHERENT:2021xmm}, following the initial 2017 observation on CsI \cite{COHERENT:2017ipa}, thereby confirming the expected \(N^2\) dependence predicted by the Standard Model. With a light yield of $\sim$4.5 PE/keVee and a $\sim$20 keVnr threshold, it provided key calibration data and demonstrated nuclear-electronic recoil discrimination through distinct scintillation time constants. 

Building on this success, the larger CENNS750 detector (610 kg LAr) \cite{COHERENT:2022nrm} is underway and employs an array of PMTs and wavelength-shifting panels that cover 61\% of the volume, targeting similar $\sim$20 keVnr thresholds. Its design features improved shielding, a cosmic-ray veto, and a continuous purification system to maintain high optical purity. Both detectors exploit the pulsed SNS beam to achieve excellent background rejection.

\textbf{CCM at LANL:} The Los Alamos Neutron Science Center (LANSC) at Los Alamos National Laboratory hosts the Lujan Center which is another prolific source of neutrinos from $\pi$-DAR, produced when an 800 MeV proton beam interacts with a tungsten target \cite{CCM:2021leg}. The proton beam is delivered at a rate of 20 Hz in a 280 ns triangular pulse from the LANSCE beamline. 

The CCM experiment operates here, employing a 10 ton cylindrical LAr scintillation detector to study CEvNS and light dark sector interactions \cite{CCM:2021yzc, CCM:2021jmk}. The cylindrical cryostat, 2.58 m in diameter and 2.25 m high, is instrumented initially with 120 PMTs and then updgraded to 200 PMTs that provide $\sim$10 keVnr threshold and precise timing resolution. Positioned 20-40 m from the tungsten target, the detector uses optimized steel and concrete shielding to suppress beam-related neutron backgrounds. The 800 MeV proton beam produces $\pi$-DAR neutrinos in 280 ns pulses at 20 Hz, enabling time-correlated searches for rare events with excellent background rejection.

\textbf{PIP2-BD at F2D2 at FNAL:} The Proton Improvement Project-II Beam Dump (PIP2-BD) is a proposed experiment at Fermilab's Facility for Dark Matter Discovery (F2D2) \cite{Aguilar-Arevalo:2023dai, Toups:2022yxs}, optimized for high-energy physics applications, including CEvNS and dark sector studies. It will utilize the 800~MeV superconducting linac from PIP-II, the first phase of Fermilab's accelerator upgrade program supporting DUNE. The PIP-II Linac accelerates 0.55~ms pulses of H$^-$ ions to 800~MeV at a 20 Hz repetition rate with a 2 mA peak current and 1.1\% duty factor, corresponding to an available beam power of up to 1.6 MW. Unlike spallation neutron facilities optimized for neutron production, PIP2-BD is designed from the ground up for HEP applications, providing enhanced sensitivity and flexibility to host multiple $\mathcal{O}(100)$ ton detectors at varying baselines and angles relative to the beam dump.

The reference detector concept for PIP2-BD is a 100 ton cylindrical LAr scintillation detector (5 m height, 2.5 m radius) enclosed within a 6 m $\times$ 6 m $\times$ 6 m active veto cryostat. The inner LAr volume is lined with TPB coated Teflon panels and instrumented with 1,294 eight inch PMTs to achieve uniform light collection and keV scale energy sensitivity. This configuration offers a versatile platform for precision CEvNS measurements and dark sector searches, complementing the existing CEvNS program at SNS and LANSCE with comparable beam power and optimized detector geometry.


\section{Analyzing Standard and Beyond the Standard Model Signals}\label{sec:analysis}
%

The precise measurement of CEvNS recoil spectra provides an opportunity to probe both the SM parameters and potential signatures of new physics. Deviations in the total event rates or recoil energy distributions can reveal the presence of new physics beyond the SM, such as electromagnetic properties of neutrinos or non-standard neutrino interactions. In this section, we present the framework used to simulate CEvNS signals for argon-based detectors and describe the statistical procedure employed to evaluate sensitivities to these effects.  

The differential event rate as a function of nuclear recoil energy is given as,
\begin{eqnarray}\label{eq:evnt_rate}
    \frac{dN}{dT} = \frac{f_{\nu/p}\,m_T}{A\,m_N} \sum_{i = \nu_\mu, \nu_e, \bar\nu_\mu} \int_{E_\nu^{min}}^{m_\mu/2} dE_\nu \frac{d\phi_{i}}{dE_\nu}\left[\frac{d\sigma (E_\nu)}{dT}\right]_{\rm SM}
\end{eqnarray}
where $f_{\nu/p}$ is the number of neutrinos produced per proton on target, $m_T$ is the mass of the detector,  $A$ is the atomic mass number and $m_N$ is the nucleon mass. $d\phi_i/dE_\nu$ is the energy spectrum of the neutrino flavor $i$. The integration limits are set by the kinematics of pion decay at rest neutrino spectrum. The recoil spectra are computed using the fluxes and detector properties listed in Table~\ref{tab:expt_list}. The left panel of Fig. \ref{fig:Rate_Fluxes} shows the projected event rate per day for the different flux sources using Ar as the target. We follow a similar prescription as the COHERENT collaboration \cite{COHERENT:2020iec} which binned their data after including smearing effects as reconstructed electron-equivalent recoil energy ($T_{\rm rec}$). Assuming a detector like CENNS10 for all flux sources, the total expected number of events in bin $i$, taking into account the detector efficiency 
$\epsilon$, and the energy resolution ($\mathcal{R}$), is given by \cite{Coloma:2022avw}:
\begin{eqnarray}
    N_i &=& \int_{T_{\rm rec,i}-\Delta T_{\rm rec}/2}^{T_{\rm rec,i}+\Delta T_{\rm rec}/2} dT_{\rm rec} \, \epsilon(T_{\rm rec})\times \nonumber \\ && \int_{T_{\text{min}}}^{\infty} dT \frac{dN}{dT} \mathcal{R}(T_{\rm rec}, T_I; \sigma_I),
\end{eqnarray}
where the energy resolution is modeled by a Gaussian distribution of width $\sigma_I = 0.58 \, \text{keV} \sqrt{T_I/\text{keV}}$, and $T_I$ denotes the true electron-equivalent recoil energy:
\begin{eqnarray}
    \mathcal{R}(T_{\text{rec}}, T_I; \sigma_I) = \frac{1}{\sqrt{2\pi} \sigma_I} e^{-\frac{(T_{\text{rec}} - T_I)^2}{2\sigma_I^2}}.
\end{eqnarray}
The true electron-equivalent recoil energy is related to the nuclear recoil energy through an energy-dependent quenching factor, parameterized as \(Q_F(T) = 0.246 + 0.00078\,T\), such that \(T_I = Q_F(T)\,T\). The resulting reconstructed event distributions are shown in the right panel of Fig. \ref{fig:Rate_Fluxes} for three years of operation. 

To evaluate sensitivity of a new physics model or the SM physics, characterized by parameters $\lambda$, we perform a statistical analysis of all experiments considered here. The analysis presented here is evaluated by considering a run time of 3 years with each year accounting for approximately 5000 hours of continual operation like that in SNS source for CENNS10 and CENNS750. Our primary aim is to explore key observables, such as the weak mixing angle $\sin^2\theta_W$, as well as to impose constraints on potential new physics effects such as the magnetic moment and charge radius of neutrino and NSI parameters.
To evaluate the experiment's sensitivity to the observables under investigation, we carried out a
  analysis, minimizing the simplest $\chi^2$ function:
\begin{align}
    \chi^2\left(\{\lambda\}\right) = \rm{min}_\eta \left\{\frac{\left[N_{\rm SM}-N_{\rm Signal}\left(\{\lambda\}\right) (1+\eta)\right]^2}{N_{\rm SM}} + \left(\frac{\eta}{\sigma_\eta}\right)^2\right\}
\end{align}
Where \(N_{\rm SM}\) is the total number of events as a function of \(T_{\rm rec}\) assuming no background and \(N_{\rm Signal}\), the total events including the parameters $\lambda$ tested here.

\noindent In this analysis, we assume negligible background in order to estimate the intrinsic sensitivity of the CEvNS signal. This corresponds to an idealized scenario; a more realistic treatment including detector-specific backgrounds is expected to reduce the sensitivity and will be considered in future work.

For the analysis presented here the systematic uncertainty $\sigma_\eta$ includes contributions from the neutrino flux normalization, quenching factor, detector efficiency, and nuclear form factor. As baseline, we assume a conservative overall systematic uncertainty of 10\% , consistent with current CEvNS measurements by COHERENT. A heavy-water Cherenkov detector has been proposed by the COHERENT collaboration as an in-situ neutrino flux monitor. Its implementation is expected to reduce the uncertainty in the absolute flux normalization -- and consequently the overall systematic uncertainty -- to about 5\% \cite{COHERENT:2021xhx}. A similar setup could be realized at F2D2. Accordingly, we include an optimistic 5\% systematic uncertainty scenario for both CENNS-750 and PIP2-BD at F2D2, corresponding to the potential in-situ improvement achievable with such a dedicated flux-monitoring detector.

\subsection{Precision Measurement of the Weak Mixing Angle}
The weak mixing angle, $\theta_W$, is a fundamental parameter of the electroweak theory that quantifies the mixing between the $SU(2)_L$ and $U(1)_Y$ gauge groups.
It governs the relative strength of neutral- and charged-current weak interactions and can be expressed as $\sin^2\theta_W = 1 - M_W^2 / M_Z^2$ at tree level.  
While it has been measured with high precision at energies near the $Z$-peak (around 100~GeV), its determination at low momentum transfer remains significantly less precise \cite{ParticleDataGroup:2024cfk}. A precise low-energy measurement provides a stringent test of the running of $\sin^2\theta_W(Q^2)$ predicted by the Standard Model (SM) and serves as a sensitive probe for new physics that may alter this running.
It is useful in determining a number of observable quantities, such as the ratio of the weak gauge boson masses, certain weak-interaction cross sections, and parity-violating effects. 
The CEvNS process offers an especially clean method for measuring $\sin^2\theta_W$ at low $Q^2$. Since the CEvNS cross section and thus the event rate depends on the weak mixing angle (see Eqs. \ref{eq:SM_cross_section} and \ref{eq:evnt_rate}), the observed CEvNS event counts can be used to extract information about $\theta_W$. Any shift in $\theta_W$ will lead to a corresponding change in the event rate and therefore could be a clear indication of new physics. 

\begin{figure*}
	\centering
	\includegraphics[width=0.39\linewidth]{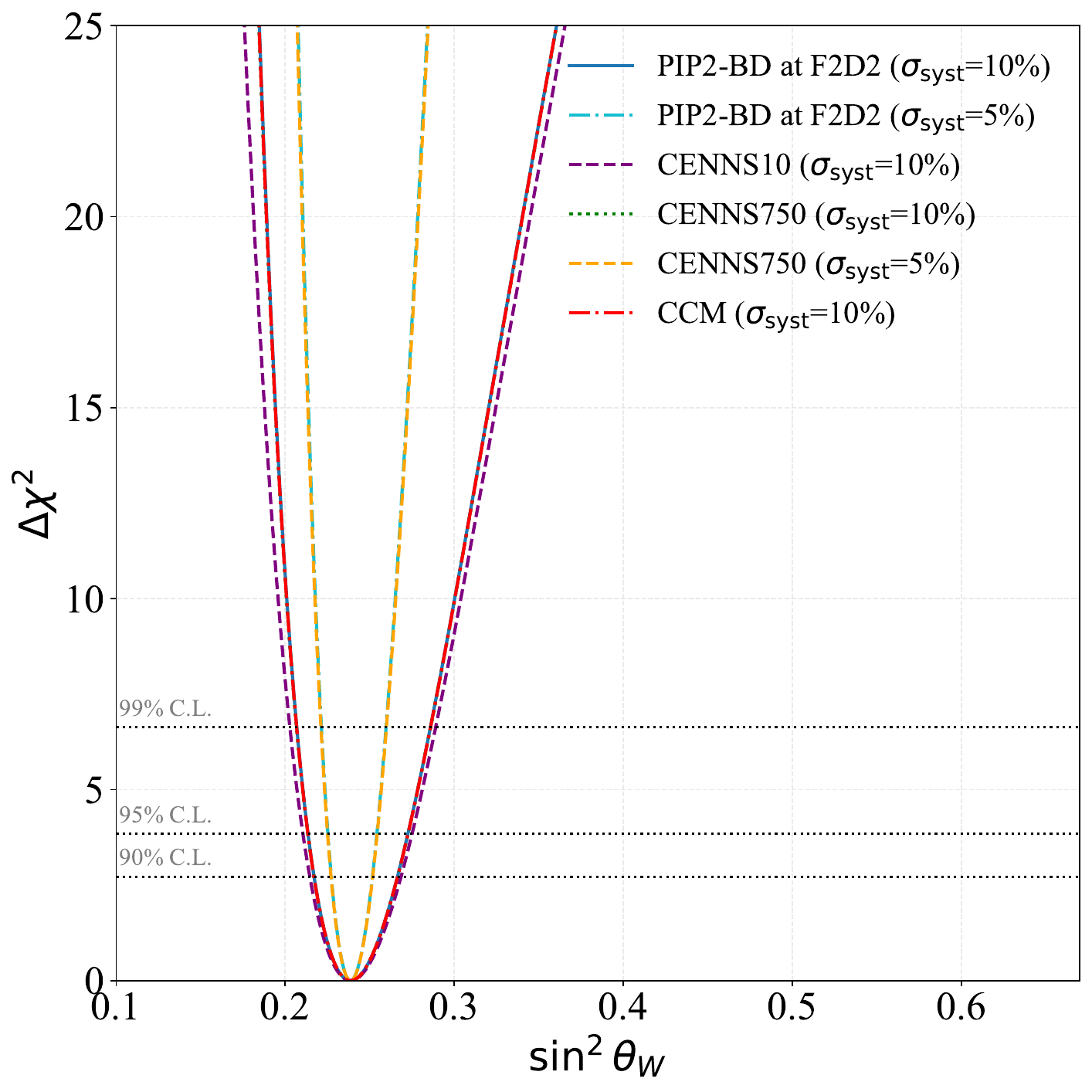}
    \includegraphics[width=0.43\linewidth]{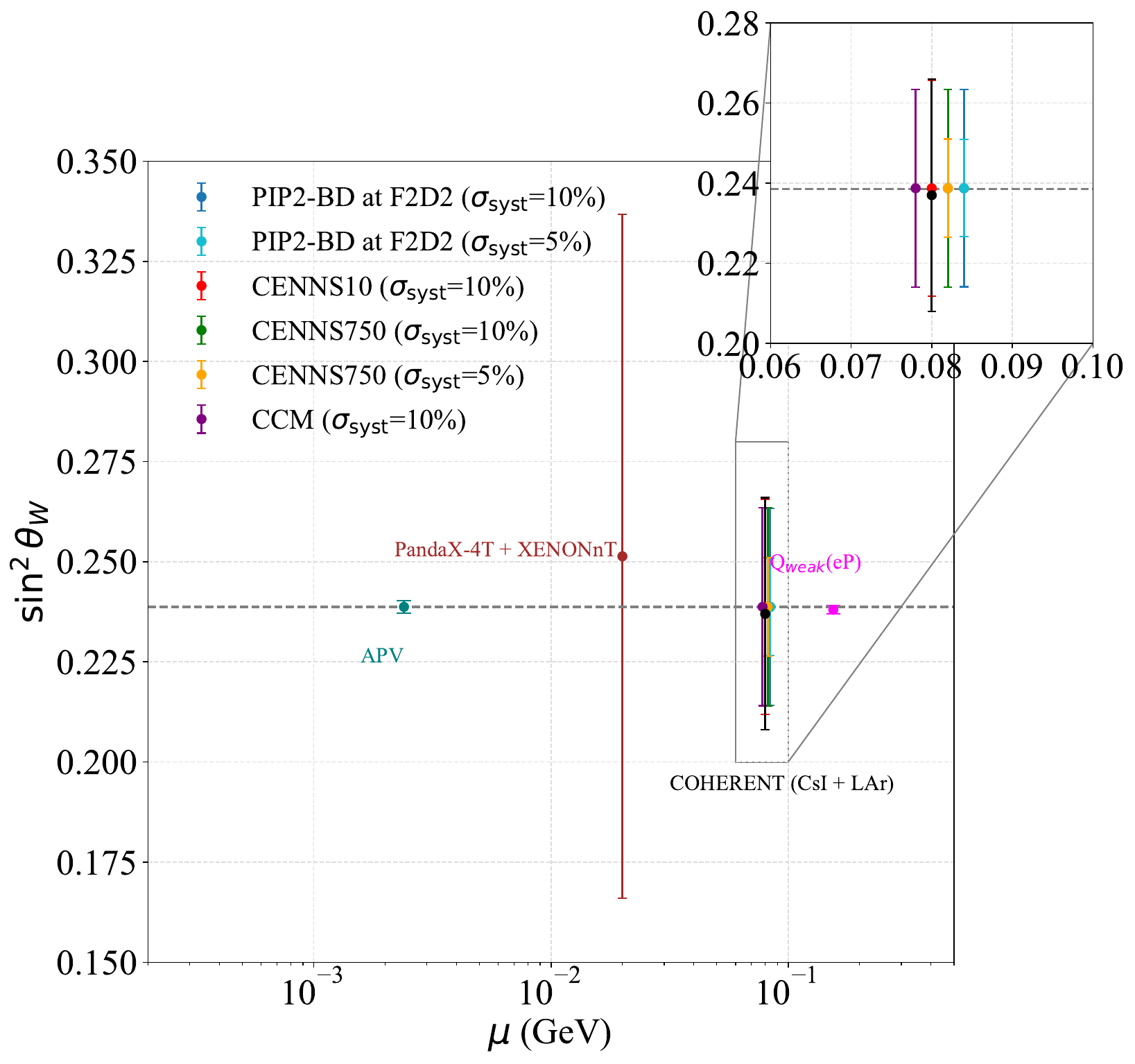}
	\caption{(Left) Sensitivity on $\sin^2\theta_W$ showing 90\%, 95\%, and 99\%~C.L. regions for different LAr experiments and systematic assumptions. (Right) Running of \(\sin^2\theta_W\) with energy scale \(\mu\) shown with prediction for different LAr experiments and systematic assumptions considered in this work. The SM prediction is shown as dashed black line alongside existing experimental measurements from Atomic Parity Violation (APV) of Cs (teal), $Q_{\mathrm weak}$ from electron-proton scattering (pink), combined analysis of PandaX-4T with XENONnT (brown) and COHERENT - CsI + LAr (black). The inset provides a magnified view of \(\theta_W\) and associated ranges from the different flux sources, with a small offset applied for readability.}
	\label{fig:chi2_sinthw}
\end{figure*}

\begin{table*}
\centering
\begin{tabular}{@{}lc@{}}
\toprule
\textbf{Experiment (Syst.\ Unc.)} & \boldmath$\sin^{2}\theta_W$ \\
\hline
PIP2-BD at F2D2 (10\%)     &$ [0.2171, 0.2663]$ \\
PIP2-BD at F2D2 (5\%)      & $[0.2273, 0.2515]$ \\
CENNS-10 (10\%)              &$ [0.2146, 0.2684]$ \\
CENNS-750 (10\%)           & $[0.2170, 0.2664]$ \\
CENNS-750 (5\%)            & $[0.2271, 0.2517]$ \\
CCM (10\%)                 & $[0.2170, 0.2664]$ \\ 
\hline
\end{tabular}
\caption{Estimated range of the weak mixing angle with respect to its central value of \(\ \sin^2\theta_W = 0.2387\) at 90\% C.L.\ for different argon-based
CEvNS experiments, assuming different levels of systematic
uncertainty. }
\label{tab:sin2theta}
\end{table*}

In Fig. \ref{fig:chi2_sinthw} (left), we show the sensitivity of $\sin^2\theta_W$ using simulated CEvNS event rates for different LAr experiments and for different systematic uncertainty assumptions. The right panel of Fig. \ref{fig:chi2_sinthw} illustrates the evolution (or ``running'') of $\sin^2\theta_W(Q^2)$ with momentum transfer \(\mu\)(in GeV), comparing our projections with existing experimental data from COHERENT (CsI + LAr)~\cite{DeRomeri:2022twg}, PandaX-4T + XENONnT \cite{Maity:2024aji, DeRomeri:2024iaw}. The resulting constraint on the weak mixing angle shows a notable improvement over previous measurements. Our results indicate that the PIP2-BD at F2D2 with reduced systematic uncertainty of 5\%, could determine the weak mixing angle with a precision competitive with or better than current low-energy determinations. CENNS-750 and CCM are also expected to reach comparable precision, enabling cross-checks across independent facilities with distinct systematics. Our best estimate for the weak mixing angle for the different experiments is listed in Table \ref{tab:sin2theta}. In the broader context, precision CEvNS measurements on these LAr experiments complement ongoing efforts in parity-violating electron scattering and neutrino-electron scattering experiments. Together, these will map the running of $\sin^2\theta_W$ over several orders of magnitude in $\mu$, providing one of the most stringent low-energy tests of the electroweak theory to date.
Any observed deviation from the SM expectation could point to new phenomena beyond the Standard Model.



\subsection{Probing Neutrino Electromagnetic Properties}
While in the SM neutrinos are massless and have interactions only in the weak sector, the discovery of neutrino oscillations confirms that at least two neutrino mass eigenstates are nonzero. The extensions of the SM to accommodate the non-zero neutrino masses can, in some scenarios, also lead to neutrinos acquiring electromagnetic properties through quantum loop effects \cite{Giunti:2014ixa}. The observation of a neutrino mass would point to a new physical scale ($\Lambda$) if the neutrinos were Majorana in nature and therefore requiring the detection of neutrinoless double beta decay. But unlike the probe of mass, the presence of neutrino interactions in the EM sector would clearly point to the existence of new physics at a higher scale. In this regime, CEvNS is particularly sensitive to such effects because the EM contribution enhances the scattering cross section at low recoil energies, leading to a distinct distortion of the nuclear recoil spectrum relative to the SM prediction.

For energies well below the electroweak scale ($\mathcal{O}(\rm MeV)$ for CEvNS), the interaction of neutrinos with photons can be described by a sum of higher-dimensional operators \cite{Altmannshofer:2018xyo, Bansal:2022zpi} up to dimension-6 as
\begin{eqnarray}\label{eq:EM_Lag}
    \mathcal{L}_{\rm EM} &\supset& \sum_{\alpha, \beta}\frac{\mathcal{C_{\rm M,\alpha\beta}^{\rm(5)}}}{\Lambda}\frac{e}{8\pi^2}(\bar\nu_\alpha\sigma^{\mu\nu}P_L\nu_\beta)F_{\mu\nu} \nonumber\\ &+& \frac{\mathcal{C_{\rm A,\alpha\beta}^{\rm(6)}}}{\Lambda^2}(\bar\nu_\alpha \gamma^\mu \gamma_5 \nu_\beta)\partial^\nu F_{\mu\nu}+ \dots
\end{eqnarray}
where $\alpha,\beta=e,\mu,\tau$ are the neutrino flavors, $\mathcal{C}^{\rm (5)}_{\rm M, \alpha \beta}, \mathcal{C}^{\rm (6)}_{\rm A, \alpha \beta}$ are the dimensionless Wilson coefficients which together with the new-physics scale describe magnetic and anapole (charge-radius) moments, respectively. In the following subsections, we examine the sensitivity of these experiments to two key electromagnetic observables: the neutrino magnetic moment and the charge radius. We quantify how each affects the CEvNS cross section and evaluate the resulting bounds achievable with current and forthcoming argon-based detectors.



\subsubsection{Neutrino Magnetic Moment}
For a Dirac neutrino, with the inclusion of a right-handed counterpart in the SM, the mass term and the magnetic moment operator (first term in Eq. \ref{eq:EM_Lag}) have the same chiral structure. This results in the magnetic moment being proportional to the Dirac mass term (\(m_\nu\)) with the value given in terms of Bohr Magneton (\(\mu_B\)) as \cite{Fujikawa:1980yx},
\[\mu_\nu = 3 \times 10^{-19}\,\mu_{\text{B}} \left(\frac{m_\nu}{\text{eV}}\right).\]
In contrast, for Majorana neutrinos, the (transition) magnetic moment \cite{Pal:1981rm} is three orders of magnitude smaller than that of Dirac neutrinos. These values of magnetic moment for neutrino mass of \(\mathcal{O}\)(eV), are many orders of magnitude smaller than the current experimental limits. The neutrino magnetic moment can be significantly enhanced, to varying degrees, in theories involving new physics at the TeV scale. Some models, such as the left-right symmetric model and supersymmetric extensions of the SM \cite{Giunti:2014ixa}, require fine-tuning of parameters to obtain a realistic value for the neutrino mass. However, even with such enhancements, the predicted magnetic moment typically remains below current experimental sensitivity. Larger enhancements, approaching the experimental limits, can be achieved by extending left-right symmetric models with an SU(2)$_\text{L}$ singlet charged scalar \cite{Fukugita:1987ti, Babu:1987be, Rajpoot:1990hj}. These scenarios also involve enhancement in the neutrino mass beyond acceptable limits unless the models are very precisely fine-tuned. The suppression of the neutrino mass alongside an enhanced magnetic moment can be achieved through the \textit{Voloshin mechanism} \cite{Voloshin:1987qy}. By exploiting the distinct Lorentz structures of the mass and magnetic moment operators, a new symmetry such as SU(2)$_{\rm H}$ \cite{Babu:2020ivd} can be introduced to nullify the mass term while permitting a magnetic moment of the order of current experimental limits.

\begin{figure}
	\centering
	\includegraphics[width=0.90\linewidth]{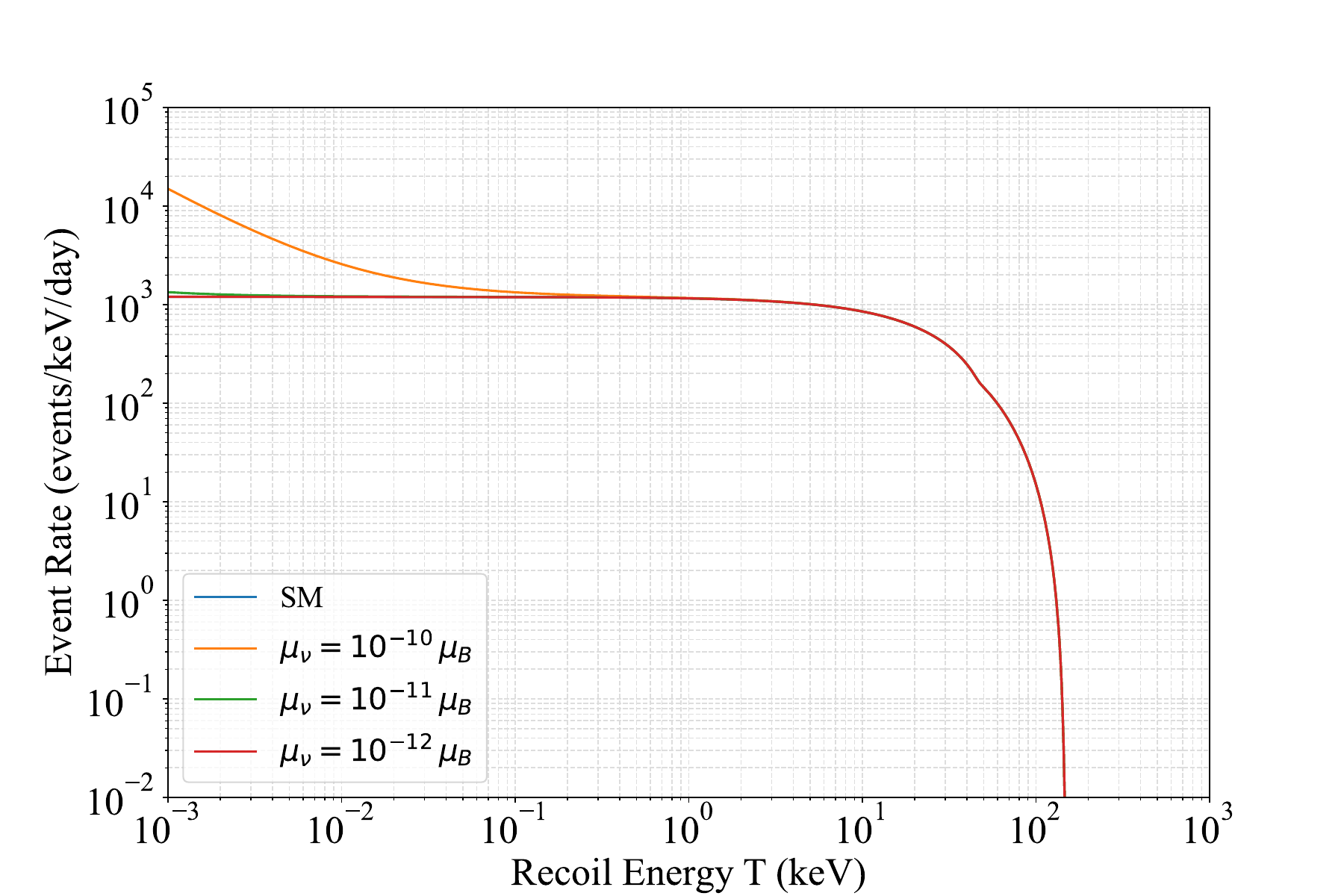}
	\caption{CEvNS event rate per day (for PIP2-BD at F2D2) as a function of the nuclear recoil energy due to Neutrino Magnetic moment for three different values for the moment (\(\mu_\nu = 10^{-10},\,10^{-11}\,\)and \(10^{-12}\,\mu_B\)).}
	\label{fig:Rate_MM}    
\end{figure}

\begin{figure*}
	\centering
	\includegraphics[width=0.4\linewidth]{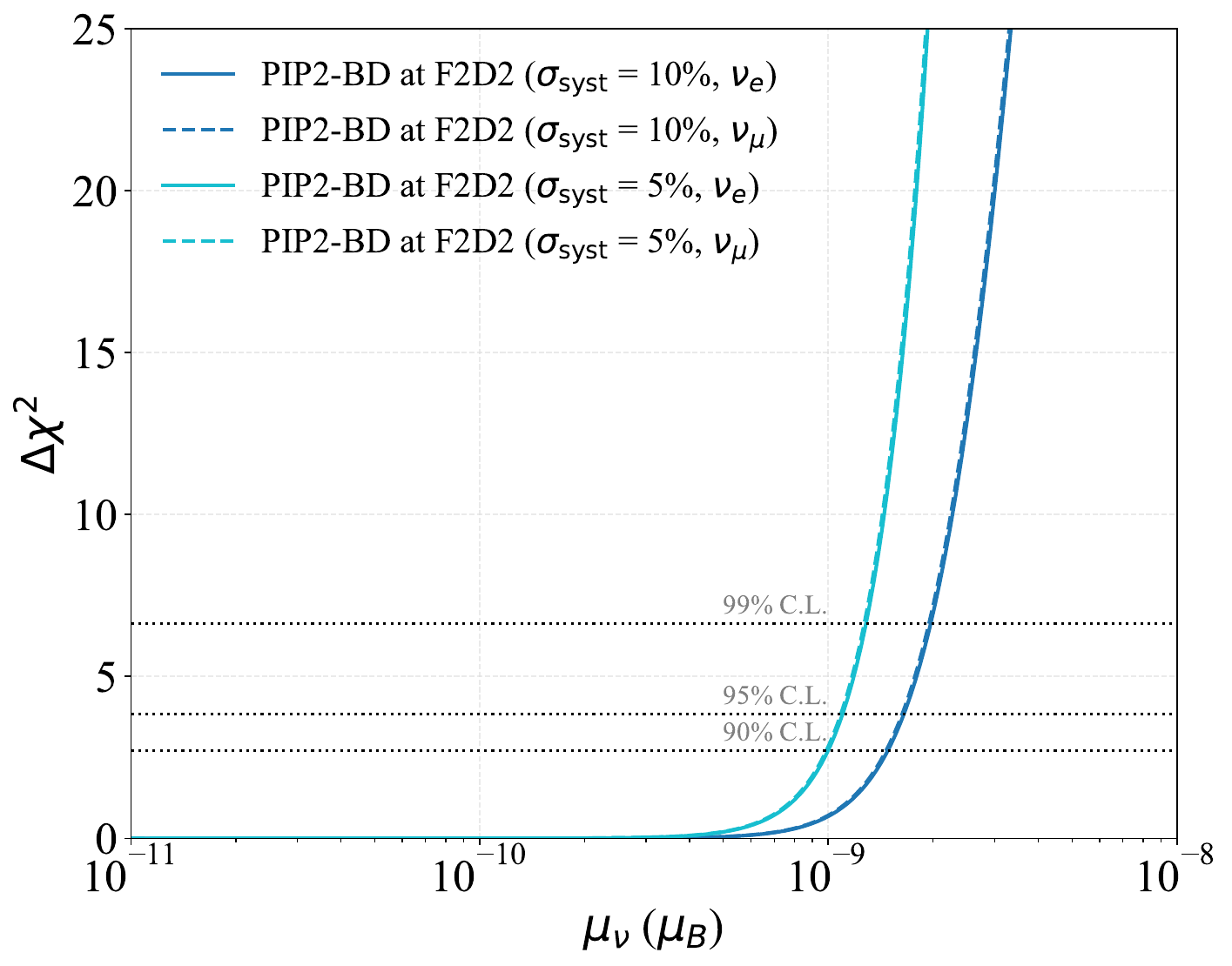}
    \includegraphics[width=0.4\linewidth]{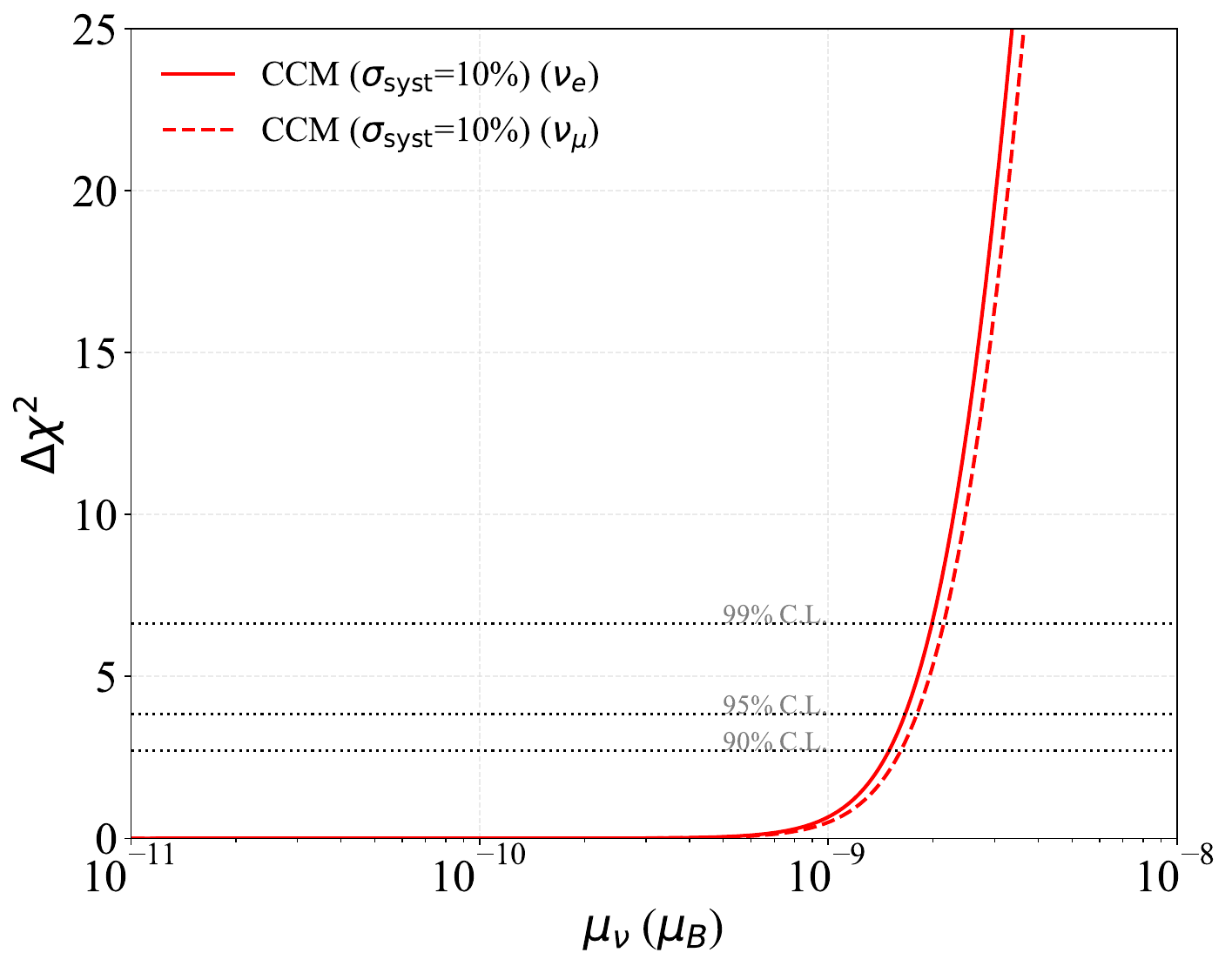}
    \includegraphics[width=0.4\linewidth]{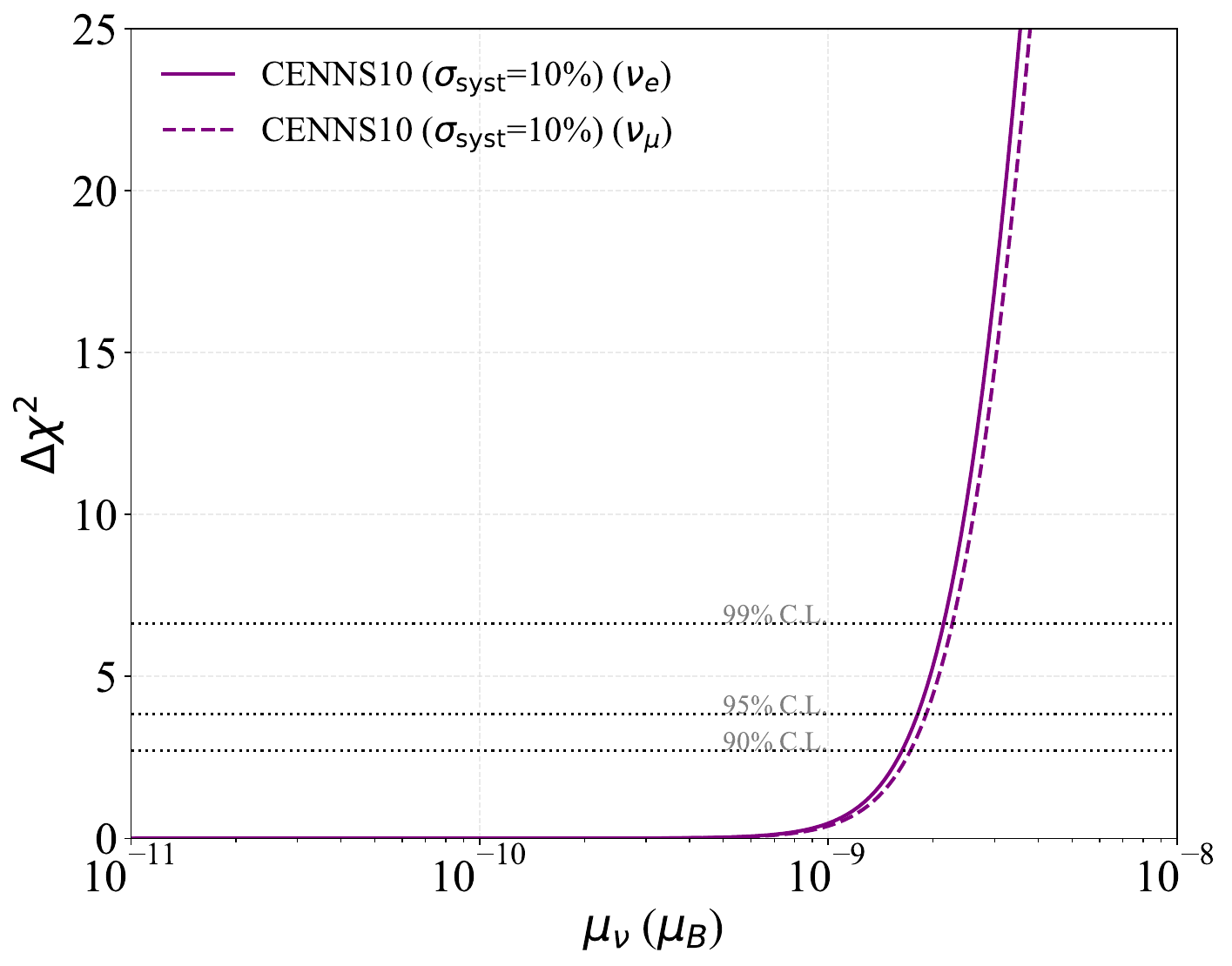}
    \includegraphics[width=0.4\linewidth]{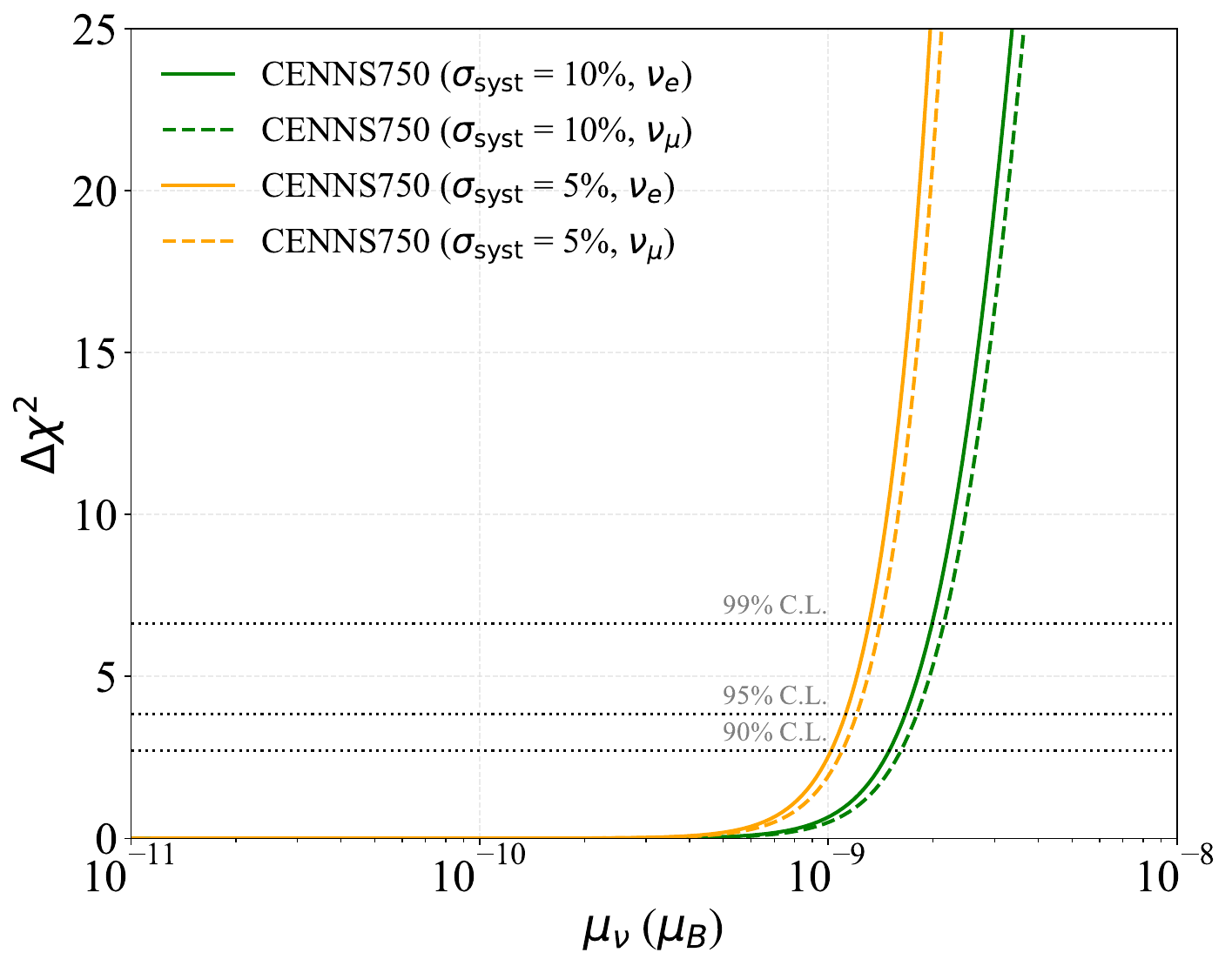}
	\caption{Sensitivity (\(\Delta \chi^2 = \chi^2-\chi^2_{\rm min}\)) on neutrino magnetic moment (\(\mu_{\nu_e},\,\mu_{\nu_\mu}\)) for different experiments: PIP2-BD at F2D2 with \(\sigma_{\rm{syst}}\) = 10\% and 5\% (top left), CCM with \(\sigma_{\rm{syst}}\) = 10\% (top right), CENNS10 with \(\sigma_{\rm{syst}}\) = 10\% (bottom left), and CENNS750 with \(\sigma_{\rm{syst}}\) = 10\% and 5\% (bottom right).}
	\label{fig:MM_Flvr_all_Sources}
\end{figure*}

In the presence of the neutrino magnetic moment, the differential cross section due to it given as,
\begin{align}
    \left[\frac{d\sigma (E_\nu)}{dT}\right]_{\rm MM} = \frac{\pi\,(\alpha_{\rm EM} \,\mu_\nu Z)^2}{m_e^2}\left(\frac{1}{T}-\frac{1}{E_\nu}+\frac{T}{4\,E_\nu^2}\right)\,F^2_{ch}(Q^2)
\end{align}
adds incoherently without interference (due to chirality flipping nature of the EM contribution) to the SM cross section in Eq. \ref{eq:SM_cross_section}. Where \(m_e\), \(\alpha_{\rm EM}\) and \(F_{\rm ch}\) are the electron mass, fine-structure constant and the charge form factor of the nucleus respectively. The change in the shape of the nuclear recoil events for the PIP2-BD at F2D2 is shown in Fig. \ref{fig:Rate_MM}. Because the electromagnetic component scales as $1/T$, it produces an enhancement at low recoil energies, leading to a distinct spectral distortion that can be experimentally separated from the SM expectation.
This characteristic feature makes CEvNS particularly sensitive to magnetic-moment effects in detectors with low thresholds and excellent background control. The current best limits on the neutrino magnetic moment come from solar and astrophysical neutrinos \cite{ParticleDataGroup:2024cfk}. The strongest (indirect) constraint (at 90\% CL \(\mu_\nu<4.5\times 10^{-12}\mu_{\rm B}\)) comes from observations of the globular cluster M5 \cite{Viaux:2013lha}. Neutrino-electron scattering provides the strongest (direct) constraint (at 90\% CL \(\mu_\nu<6.4\times 10^{-12}\mu_{\rm B}\)) based on the electronic recoil data from XENONnT \cite{XENON:2022ltv}. 

\begin{figure*}
	\centering
	\includegraphics[width=0.42\linewidth]{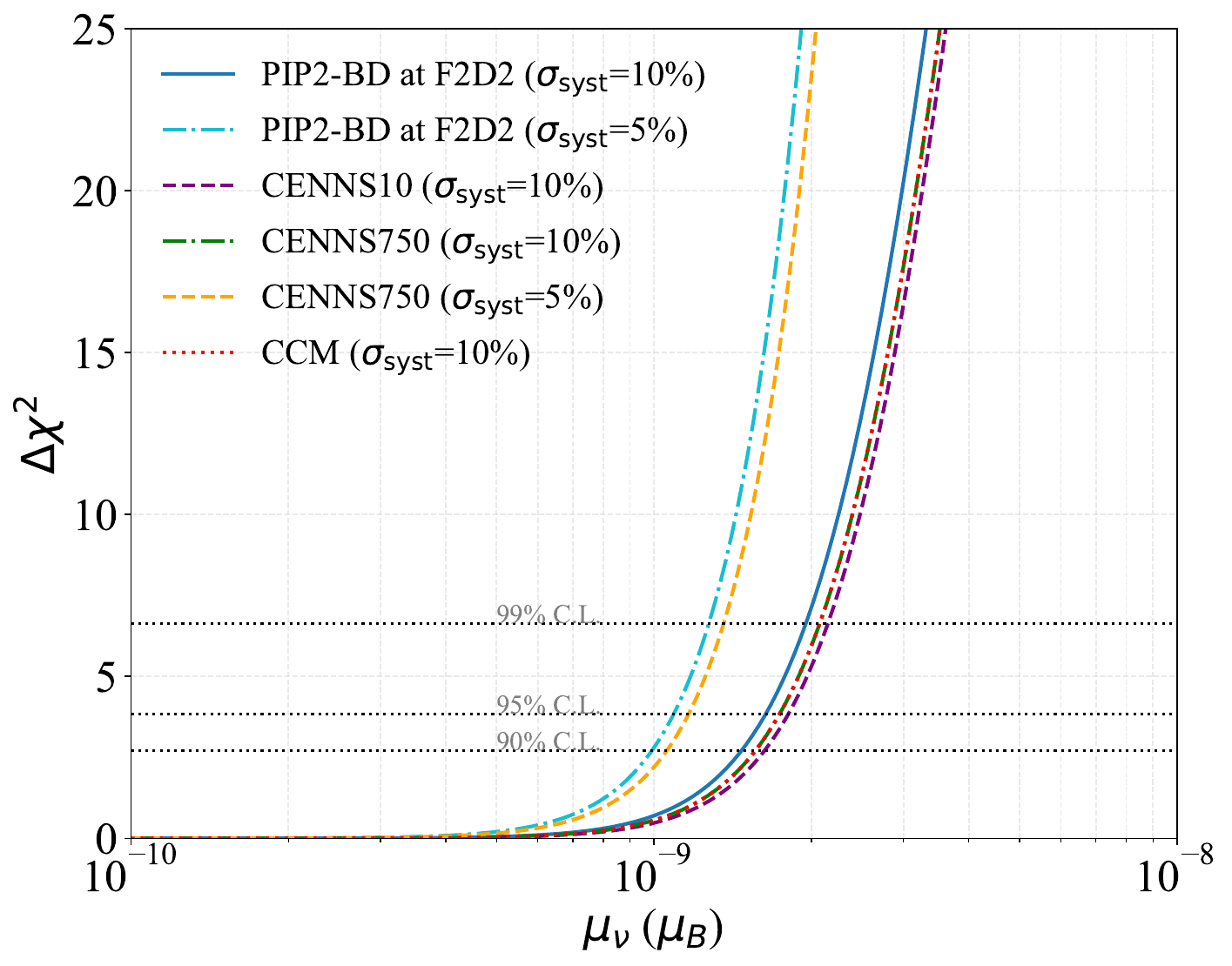}
    \includegraphics[width=0.42\linewidth]{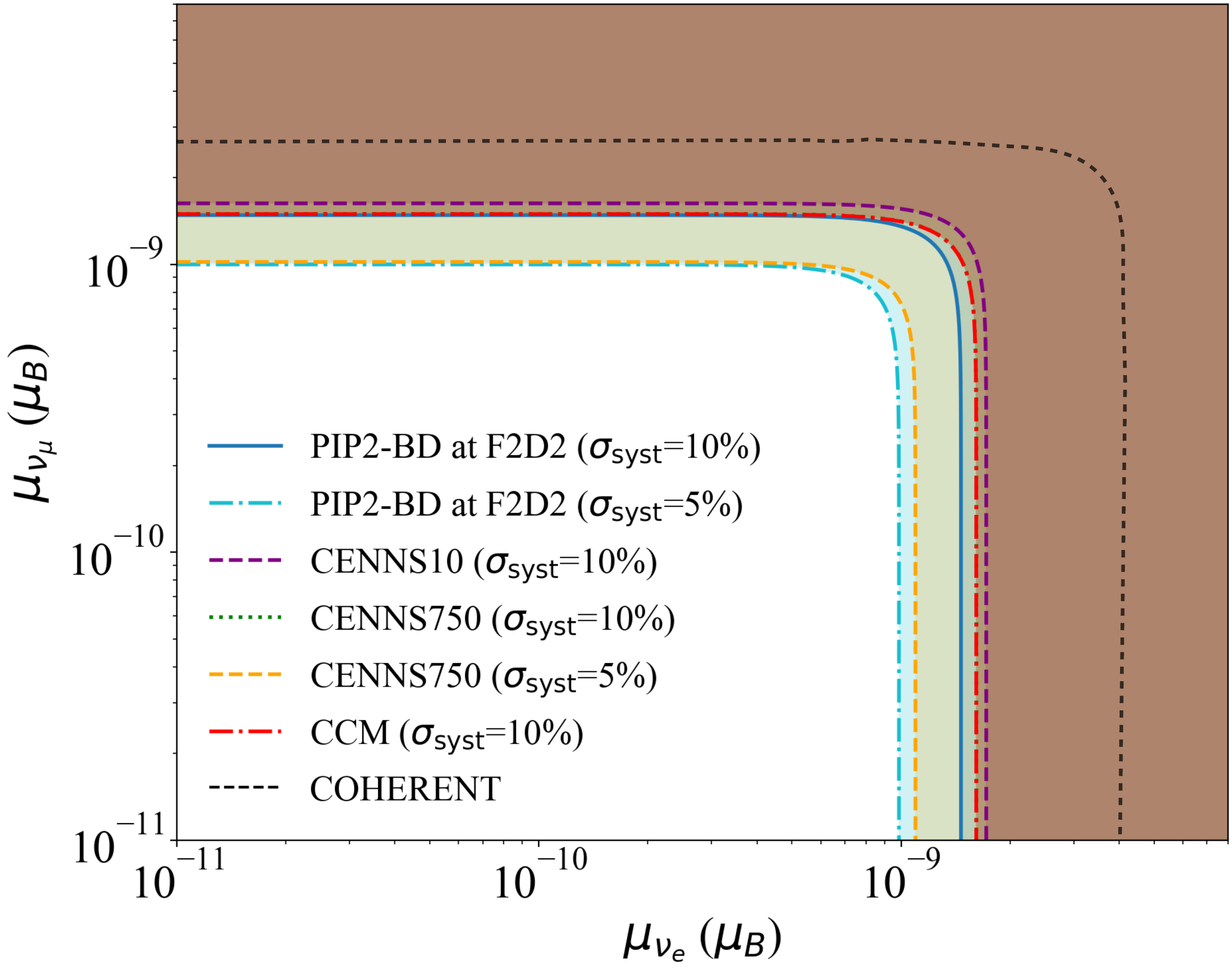}
	\caption{(Left) Sensitivity (\(\Delta \chi^2 = \chi^2-\chi^2_{\rm min}\)) on universal effective neutrino magnetic moment for the different experiments. (Right) 90\%~C.L. exclusion regions in the ($\mu_{\nu_e}$, $\mu_{\nu_\mu}$) parameter space for the different experiments with the respective systematics is shown in color. The black dotted line is the bound on these parameters from COHERENT measurements using LAr \cite{COHERENT:2020iec, Miranda:2020tif}.}
	\label{fig:MM_Cmbnd_Flvr}
\end{figure*}

In Fig. \ref{fig:MM_Flvr_all_Sources} and the left panel of Fig. \ref{fig:MM_Cmbnd_Flvr} we show the sensitivity of the flavor dependent  (\(\mu_{\nu_e},\,\mu_{\nu_\mu}\)) and universal effective neutrino magnetic moment using the simulated CEvNS event rates for the various flux sources. Our best estimate for the magnetic moment for the different flux sources is listed in Table \ref{tab:mag_moment}. For the proposed PIP2-BD configuration with a 5\% flux uncertainty, the expected upper limit on the neutrino magnetic moment reaches improves upon current CEvNS bounds by nearly an order of magnitude. This sensitivity is competitive with recent limits and approaches the region predicted by several BSM scenarios.

\begin{table*}
\centering
\begin{tabular}{@{}lccc@{}}
\toprule
\textbf{Experiment (Syst.\ Unc.)} & \boldmath$\mu_\nu $  & \boldmath$\mu_{\nu_e} $ & \boldmath$\mu_{\nu_\mu}$ \\
\hline
PIP2-BD at F2D2 (10\%)     & $14.7$ & $14.8$ & $14.7$ \\
PIP2-BD at F2D2 (5\%)      & $9.93$& $10.0$& $9.9$\\
CENNS-10 (10\%)            & $16.3$ & $16.3$&$17.2$\\
CENNS-750 (10\%)          & $15.6$ &$15.0$ &$16.2$\\
CENNS-750 (5\%)           & $10.6$ &$10.2$ &$11.0$\\
CCM (10\%)           & $15.6$ &$15.0$ &$16.2$\\
\hline
\end{tabular}
\caption{Estimated values of the neutrino magnetic moment (\(\times\, 10^{-10}\mu_{\text{B}}\)) at 90\% C.L.\ for different argon-based CEvNS experiments, assuming different levels of systematic uncertainty.}
\label{tab:mag_moment}
\end{table*}

In the right panel of Fig. \ref{fig:MM_Cmbnd_Flvr}, for \(\nu_e,\,\nu_\mu\) we show the excluded region at 90\% CL for the different flux sources under consideration with the respective systematic uncertainty. The sensitivity for \(\mu_{\nu_\mu}\) is stronger than the COHERENT measurement \cite{COHERENT:2020iec} and the joint fit with Dresden-II \cite{Colaresi:2021kus} (where the scattering off electron is included as well). But for \(\mu_{\nu_e}\) it is weaker than the joint fit while better in comparison to COHERENT alone.    

\subsubsection{Neutrino Charge Radius}
In addition to the magnetic moment, massive neutrinos may also possess an effective charge radius, $\langle r^2\rangle_\nu$, which encapsulates loop-induced effects to the neutrino-photon vertex. The effective charge radius modifies the neutral-current coupling of neutrinos to quarks and electrons and thus provides another sensitive probe of new physics.
Neutrino-photon couplings are captured by an effective electromagnetic vertex given by
\begin{align}
    \langle \nu_f |J_\mu^{\rm EM}|\nu_i\rangle = \bar u_f \left[F_1(q^2)\gamma_\mu+...\right]u_i 
\end{align}
where $F_1$ is the electric form factor (here we have ignored the other form factors for the convenience of our discussion here). Even if the electric charge is zero, that is, $F_1(0) = 0$, the neutrino charge radius squared is then defined as \cite{Giunti:2014ixa},
\begin{align}
    \langle r^2 \rangle_\nu = 6 F_1^\prime(q^2)\Big |_{q^2 = 0} 
\end{align}
The origin of the neutrino charge radius can also be understood through the dimension-6 anapole moment operator (second term in Eq.~\ref{eq:EM_Lag}). Although the anapole moment originally proposed by Zeldovich \cite{Zeldovich:1958} has a distinct physical interpretation as a parity-violating interaction, it is phenomenologically equivalent to the charge radius in the ultra-relativistic limit of the neutrino. 

The presence of the neutrino charge radius can be expressed, at leading order, as an effective shift in the weak mixing angle~\cite{Degrassi:1989ip, Vogel:1989iv},
\begin{align}
     \sin^2\theta_W \rightarrow \sin^2\theta_W \left(1 + \frac{m_W^2 \langle r^2 \rangle_{\nu_\alpha}}{3}\right),
\end{align}
where \(m_W\) is the mass of the $W$ boson. 

In this approximation, a positive (negative) value of $\langle r^2\rangle_\nu$ effectively increases (decreases) the weak charge $Q_W$ defined in Eq.~(\ref{eq:weak_charge}), leading primarily to a change in the overall normalization of the CEvNS event rate.

More generally, the neutrino charge radius arises from a momentum-dependent correction to the neutrino electromagnetic form factor proportional to $q^2 \langle r^2 \rangle_{\nu_\alpha}$, where in CEvNS kinematics $q^2 \simeq -2 m_N T$. In the low-$q^2$ regime relevant for CEvNS, this dependence can be largely absorbed into the effective shift of $\sin^2\theta_W$, with only sub-leading effects inducing mild distortions of the recoil spectrum.

\begin{table*}
\centering
\begin{tabular}{@{}lccc@{}}
\toprule
\textbf{Experiment (Syst.\ Unc.)} & \boldmath$\langle r^2 \rangle_{\nu}$ & \boldmath$\langle r^2 \rangle_{\nu_e}$ & \boldmath$\langle r^2 \rangle_{\nu_\mu}$ \\
\hline
PIP2-BD at F2D2 (10\%)   & $[-13.10, -12.06]\,\cup\, [-0.44, 0.58]$  & $[-13.08, -12.08]\,\cup\, [-0.44, 0.56]$ &$[-13.08, -12.08]\,\cup\, [-0.44, 0.56]$ \\
PIP2-BD at F2D2 (5\%)   & $[-12.78, -12.28]\,\cup\, [-0.24, 0.26]$   & $[-12.76, -12.28]\,\cup\, [-0.24, 0.24]$ &$[-12.76, -12.28]\,\cup\, [-0.24, 0.24]$ \\
CENNS-10 (10\%)          & $[-13.14, -12.02]\,\cup\, [-0.50, 0.62]$         & $[-13.16, -12.00]\,\cup\, [-0.52, 0.64]$ &$[-13.16, -12.00]\,\cup\, [-0.52, 0.64]$ \\
CENNS-750 (10\%)      & $[-13.10, -12.06]\,\cup\, [-0.46, 0.58]$     & $[-13.08, -12.08]\,\cup\, [-0.44, 0.56]$ &$[-13.08, -12.08]\,\cup\, [-0.44, 0.56]$ \\
CENNS-750 (5\%)       & $[-12.78, -12.28]\,\cup\, [-0.24, 0.26]$     & $[-12.80, -12.28]\,\cup\, [-0.24, 0.28]$ &$[-12.76, -12.28]\,\cup\, [-0.24, 0.24]$ \\
CCM (10\%)      & $[-13.10, -12.06]\,\cup\, [-0.46, 0.58]$      & $[-13.08, -12.08]\,\cup\, [-0.44, 0.56]$ &$[-13.08, -12.08]\,\cup\, [-0.44, 0.56]$ \\
\hline
\end{tabular}
\caption{Estimated two possible range of the neutrino charge radius - \(\langle r^2 \rangle_{\nu}, \,\langle r^2 \rangle_{\nu_e}, \,\langle r^2 \rangle_{\nu_\mu}\) \((\times\, 10^{-32}\text{cm}^2)\) at 90\% C.L.\ for various CEvNS flux sources and detector configurations, assuming different levels of systematic uncertainty.}
\label{tab:charge_radius}
\end{table*}

The strongest limit on the charge radius (\([-2.1, 3.3]\times 10^{-32} \rm cm^2\)) is obtained from neutrino-electron scattering by TEXONO~\cite{TEXONO:2009knm}. The best limit for coherent scattering on CsI based on data from the COHERENT collaboration is \([-27.5, 3.0]\times 10^{-32} \rm cm^2\)~\cite{Cadeddu:2018dux}. We evaluate the sensitivity to $\langle r^2\rangle_\nu$ for argon-based CEvNS experiments using the $\chi^2$ framework outlined above, assuming a three-year exposure. The corresponding one-dimensional $\Delta\chi^2$ profiles for both the flavor-dependent and universal effective neutrino charge radius are presented in Appendix~\ref{app:sensitivity} (see Figs.~\ref{fig:Charge_radius_Flvr_all_Sources} and~\ref{fig:Charge_Radius}). 
The resulting 90\% C.L. constraints are summarized in Table~\ref{tab:charge_radius}. Since the allowed regions consist of two disjoint intervals, the reported bounds correspond to the union of these ranges.

The projected bounds indicate that future CEvNS measurements with large liquid-argon detectors can reach sensitivities to the neutrino charge radius at the level of $\mathcal{O}(10^{-32})$~cm$^2$, comparable to or slightly better than the current limits.  Such precision represents an important step toward probing loop-induced electromagnetic effects in the neutrino sector. In addition, these measurements provide an independent cross-check of electroweak couplings and complement ongoing low-energy precision programs, helping to close the gap between laboratory and astrophysical probes of neutrino electromagnetic properties.

\begin{table*}
\centering
\begin{tabular}{@{}lcc@{}}
\toprule
\textbf{Experiments (Syst.\ Unc.)} & \boldmath$\epsilon_{ee}^{uV}$ & \boldmath$\epsilon_{ee}^{dV}$\\
\hline
PIP2-BD at F2D2 (10\%)  & $[-0.048,\, 0.046]\,\cup\,[0.320,\, 0.414]$ & $[-0.046,\, 0.042]\,\cup\,[0.300,\, 0.386]$\\
PIP2-BD at F2D2 (5\%)   & $[-0.024,\, 0.022]\,\cup\,[0.342,\, 0.388]$ & $[-0.022,\, 0.020]\,\cup\,[0.320,\, 0.364]$\\
CENNS-10 (10\%)         & $[-0.044,\, 0.040]\,\cup\,[0.324,\, 0.408]$ & $[-0.040,\, 0.038]\,\cup\,[0.304,\, 0.382]$\\
CENNS-750 (10\%)        & $[-0.040,\, 0.036]\,\cup\,[0.330,\, 0.406]$ & $[-0.038,\, 0.032]\,\cup\,[0.308,\, 0.378]$\\
CENNS-750 (5\%)         & $[-0.018,\, 0.018]\,\cup\,[0.348,\, 0.384]$ & $[-0.018,\, 0.016]\,\cup\,[0.326,\, 0.360]$\\
CCM (10\%)              & $[-0.040,\, 0.036]\,\cup\,[0.330,\, 0.406]$& $[-0.038,\, 0.032]\,\cup\,[0.308,\, 0.378]$ \\
\hline
\end{tabular}
\caption{Estimated two possible range for the NSI parameters \(\epsilon_{ee}^{uV}\) and \(\epsilon_{ee}^{dV}\) at 90\% C.L. for different argon-based CEvNS experiments, assuming different levels of systematic uncertainty.}
\label{tab:NSI_ee_uV_dV}
\end{table*}

\begin{table*}
\centering
\begin{tabular}{@{}lcc@{}}
\toprule
\textbf{Experiments (Syst.\ Unc.)} & \boldmath$\epsilon_{\mu\mu}^{uV}$ & \boldmath$\epsilon_{\mu\mu}^{dV}$ \\
\hline
PIP2-BD at F2D2 (10\%)  & $[-0.024,\, 0.020]\,\cup\,[0.346,\, 0.388]$ & $[-0.022,\, 0.018]\,\cup\,[0.324,\, 0.364]$\\
PIP2-BD at F2D2 (5\%)   & $[-0.010,\, 0.010]\,\cup\,[0.356,\, 0.376]$ & $[-0.010,\, 0.008]\,\cup\,[0.332,\, 0.352]$ \\
CENNS-10 (10\%)         & $[-0.028,\, 0.024]\,\cup\,[0.340,\, 0.394]$ & $[-0.026,\, 0.022]\,\cup\,[0.318,\, 0.368]$\\
CENNS-750 (10\%)        & $[-0.026,\, 0.022]\,\cup\,[0.344,\, 0.392]$ & $[-0.024,\, 0.020]\,\cup\,[0.322,\, 0.366]$\\
CENNS-750 (5\%)         & $[-0.012,\, 0.012]\,\cup\,[0.354,\, 0.378]$ & $[-0.012,\, 0.010]\,\cup\,[0.332,\, 0.352]$\\
CCM (10\%)              & $[-0.026,\, 0.022]\,\cup\,[0.344,\, 0.392]$ & $[-0.024,\, 0.020]\,\cup\,[0.322,\, 0.366]$\\
\hline
\end{tabular}
\caption{Estimated two possible range for the NSI parameters \(\epsilon_{\mu\mu}^{uV}\) and \(\epsilon_{\mu\mu}^{dV}\) at 90\% CL for different argon-based
CEvNS experiments, assuming different levels of systematic
uncertainty. }
\label{tab:NSI_mumu_uV_dV}
\end{table*}

\subsection{Sensitivity to Neutrino Non-Standard Interactions}
CEvNS provides an excellent laboratory to test for non-standard neutrino interactions (NSIs) with quarks and leptons. Such interactions naturally arise in a wide class of theories extending the SM. For energies ($\mathcal{O}(\rm MeV)$) relevant to CEvNS experiments that is well below the electroweak scale (agnostic to the nature of new physics), neutrino interaction with nucleus can be effectively described by \cite{Davidson:2003ha, Altmannshofer:2018xyo} dimension-6 operators
\begin{eqnarray}\label{eq:NSI_Lag}
    \mathcal{L}_{\rm NSI} &\supset& \sum_{\alpha, \beta,f,P}\frac{\mathcal{C_{\rm NC,\alpha\beta}^{\rm (6),f}}}{\Lambda^2}\left(\bar{\nu}_\alpha \gamma^\mu P_L \nu_\beta \right) \left(\bar{f} \gamma_\mu P f \right) \nonumber\\ &+& \frac{\mathcal{C_{\rm CC,\alpha\beta}^{\rm (6),f}}}{\Lambda^2}  \left(\bar{\nu}_\alpha \gamma^\mu P_L l_\beta\right)  \left(\bar{f}^\prime \gamma_\mu P f\right),
\end{eqnarray}
where $\alpha,\beta=e,\mu,\tau$ are the neutrino flavors, $f = u,d$ are the first generation quarks, $l = e, \mu,\tau$ are the charged leptons $P = P_L, \,P_R$ are left- and right-handed projection operators. $\mathcal{C_{\rm NC,\alpha\beta}^{\rm (6),f}}, \mathcal{C_{\rm CC,\alpha\beta}^{\rm (6),f}}$ are the dimensionless Wilson coefficients that describe the strength of neutral-current (NC) and charged-current (CC) NSIs as,
\begin{align*}
    \frac{\mathcal{C_{\rm NC,\alpha\beta}^{\rm(6), f}}}{\Lambda^2} &= 2\sqrt{2}\, G_F \,\varepsilon_{\alpha\beta}^{fV} \\
    \frac{\mathcal{C_{\rm CC,\alpha\beta}^{\rm(6), f}}}{\Lambda^2} &= 2\sqrt{2} \,G_F \,V_{ff^\prime}\,\varepsilon_{\alpha\beta}^{ff^\prime V},
\end{align*}
where \(\varepsilon_{\alpha \beta}\) are the NSI parameters and \(V_{ff^\prime}\) is the CKM matrix element. The CC NSI is strongly constrained from processes such as loop-induced muon decay and meson decay \cite{Biggio:2009nt}. Therefore, in this work here we only consider the vector type (axial type is negligible) flavor conserving ($\alpha=\beta$) NC NSI and the flavor changing ($\alpha \neq \beta$) NC NSI. In the presence of NSI, the weak charge $Q_W$ in Eq. \ref{eq:weak_charge}, is modified as \cite{Barranco:2005yy}
\begin{align}
    Q^2_{W, NSI} &= [(g_n^V + 2\varepsilon_{\alpha\alpha}^{uV} + \varepsilon_{\alpha\alpha}^{dV})N + (g_p^V + \varepsilon_{\alpha\alpha}^{uV} + 2\varepsilon_{\alpha\alpha}^{dV})Z]^2
    \nonumber \\ &+ \sum_{\alpha \neq \beta} [(\varepsilon_{\alpha\beta}^{uV} + 2\varepsilon_{\alpha\beta}^{dV})N + (2\varepsilon_{\alpha\beta}^{uV} + \varepsilon_{\alpha\beta}^{dV})Z]^2
\end{align}
The presence of NSIs leads to both rate and spectral distortions in CEvNS events. 
Because the deviation from the SM prediction depends on the combination of $Z$ and $N$ weighted by these coefficients,  
measurements with different target nuclei or at different neutrino energies can break degeneracies and constrain the allowed parameter space.
This complementarity makes CEvNS an essential component of the global NSI search program, together with oscillation and neutrino-electron scattering experiments.

In addition to scattering-based probes, neutrino oscillation experiments provide some of the most stringent constraints on flavor-dependent NSI parameters. Oscillation data are particularly sensitive to differences of diagonal couplings and to flavor-changing interactions through matter effects, leading to strong bounds on combinations of $\varepsilon_{\alpha\beta}^{fV}$, especially in the $e$-$\mu$ and $e$-$\tau$ sectors~\cite{Coloma:2016gei, Coloma:2017egw, Coloma:2023ixt}. These global analyses typically exhibit degeneracies, such as the well known generalized mass ordering degeneracy, which allow for relatively large individual NSI parameters provided specific combinations are satisfied.

In this context, CEvNS measurements play a complementary role, as they probe different linear combinations of NSI couplings weighted by the nuclear charges $(Z,N)$ and are insensitive to oscillation related degeneracies, while exhibiting their own characteristic degeneracy structures. Therefore, combining CEvNS results with oscillation data can significantly improve the determination of NSI parameters and help break degeneracies present in global fits.

\begin{figure*}
	\centering
	\includegraphics[width=0.32\linewidth]{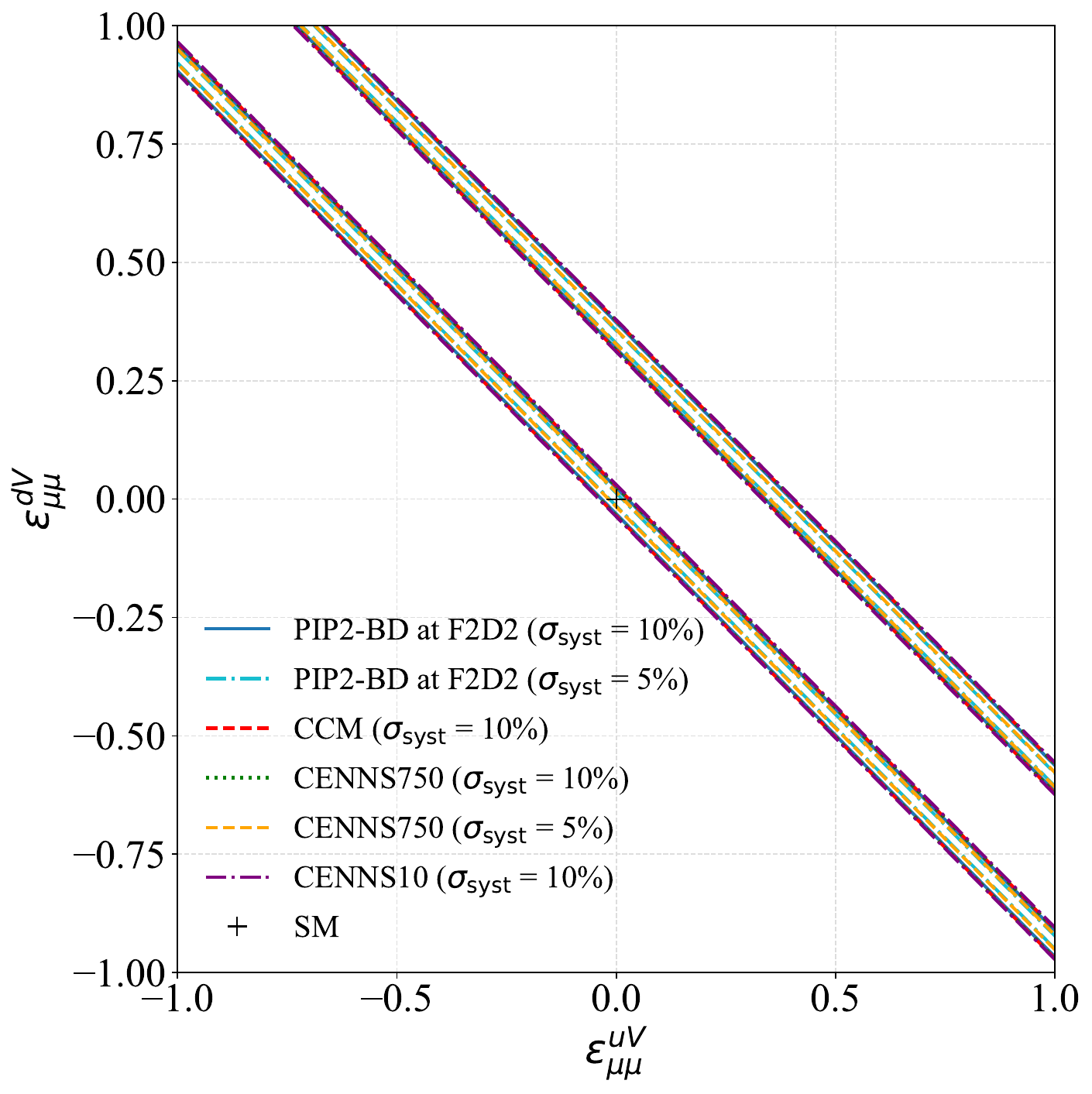}
    \includegraphics[width=0.32\linewidth]{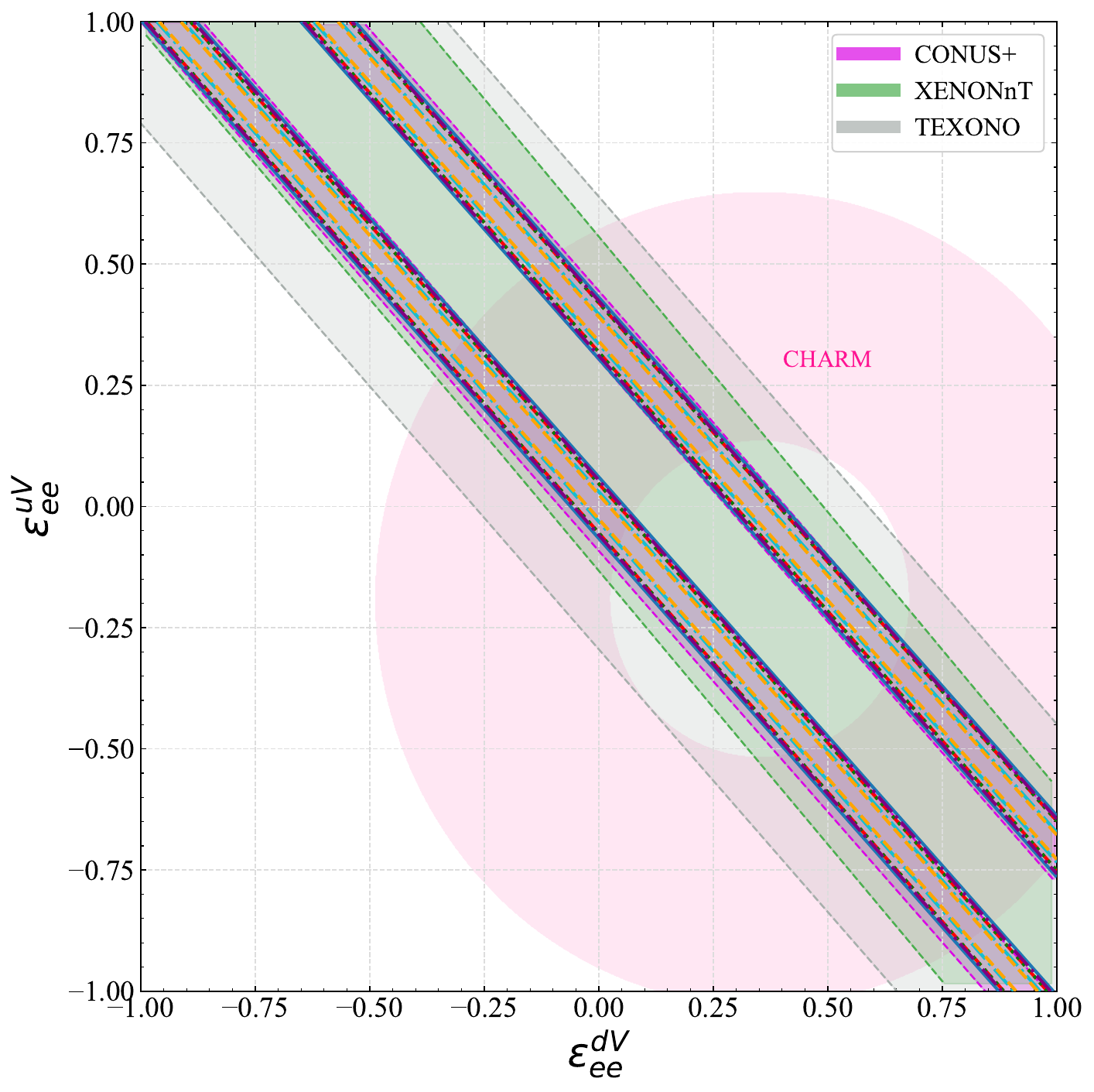}
    \includegraphics[width=0.32\linewidth]{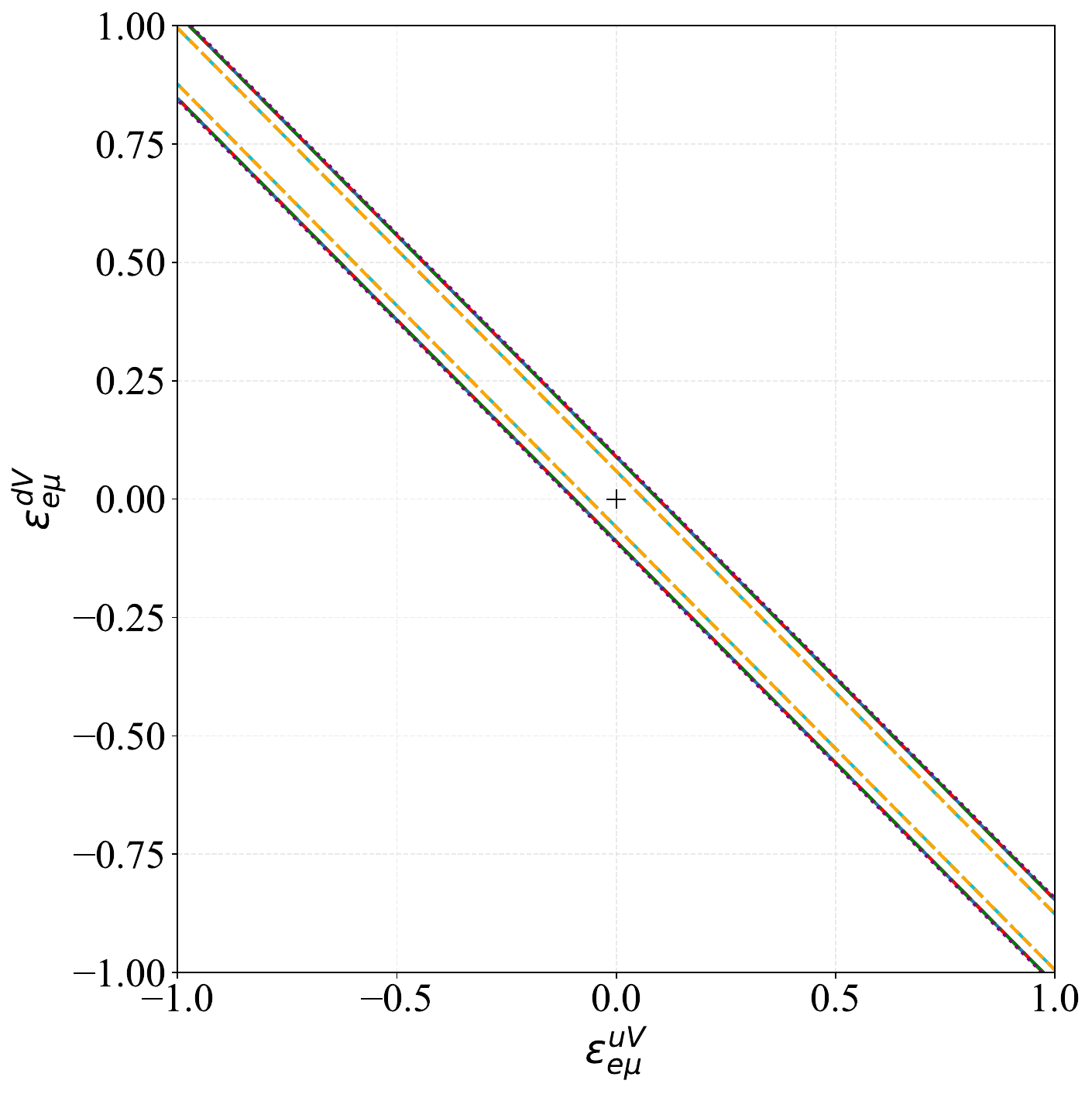}
	\caption{90\% CL allowed regions for flavor-conserving NSI parameters: \(\varepsilon^{dV}_{\mu\mu}\) vs \(\varepsilon^{uV}_{\mu\mu}\) (left) and \(\varepsilon^{dV}_{ee}\) vs \(\varepsilon^{uV}_{ee}\) (right). The results from different sources are compared with the CHARM~\cite{CHARM:1986vuz}, CONUS+~\cite{DeRomeri:2025csu}, XENONnT~\cite{AristizabalSierra:2024nwf} and TEXONA~\cite{TEXONO:2025sub} 90\% CL constraints.}
	\label{fig:NSI_Umumu_vs_Dmumu}
\end{figure*}

\begin{figure*}
	\centering
	\includegraphics[width=0.4\linewidth]{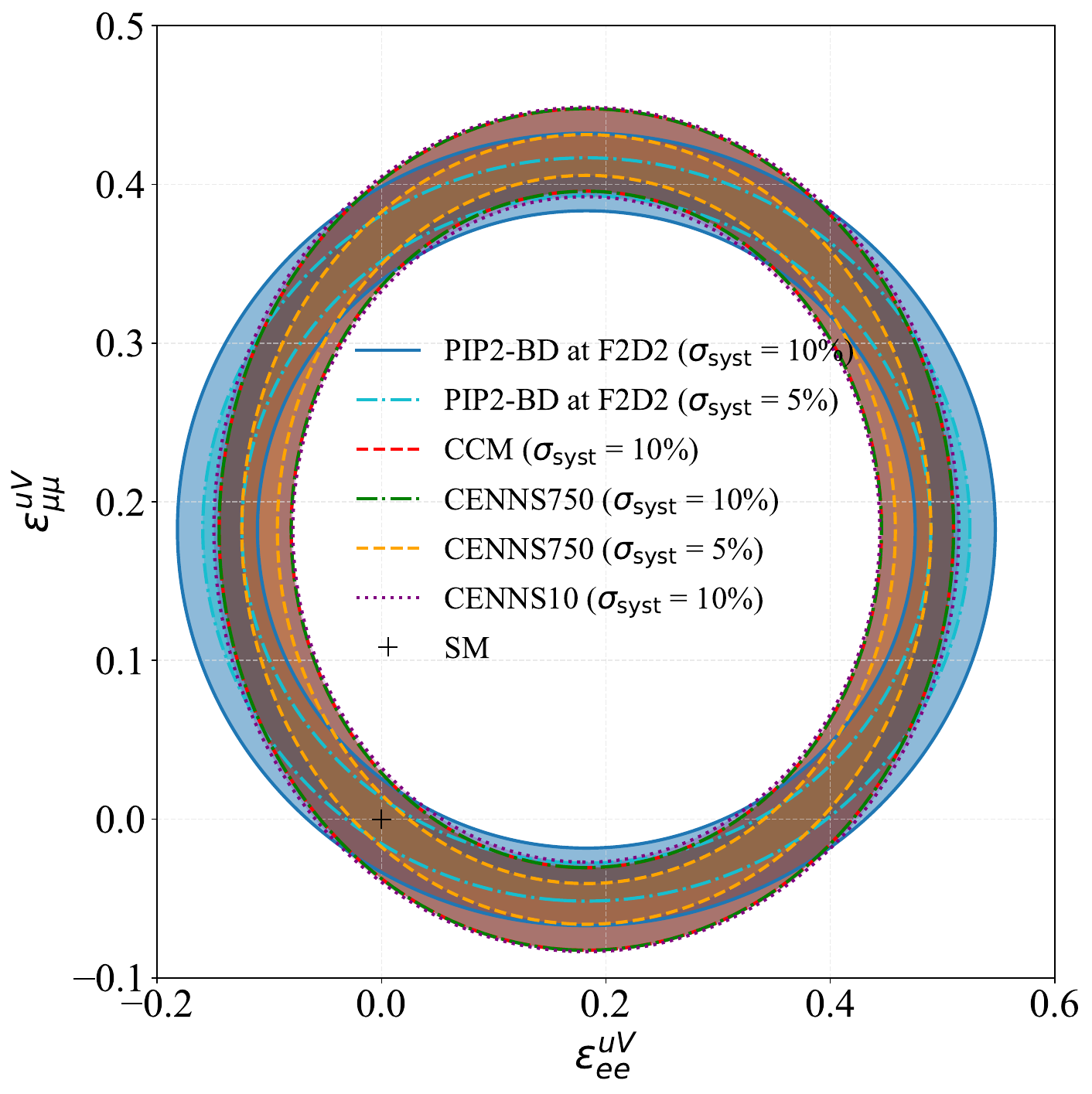}
    \includegraphics[width=0.4\linewidth]{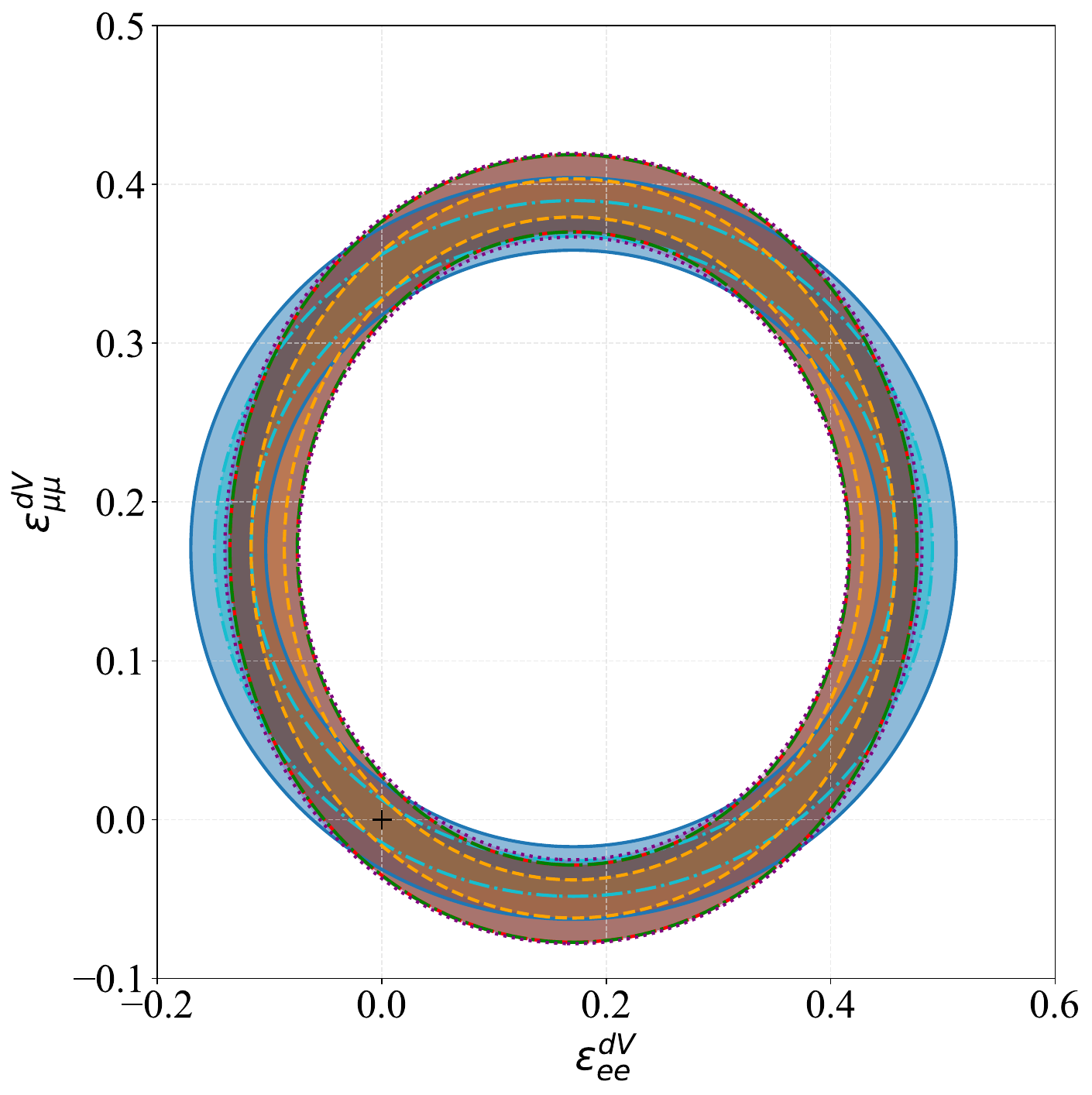}
	\caption{90\% CL allowed regions for flavor diagonal NSI parameters: (left) \(\varepsilon^{uV}_{ee}\) vs \(\varepsilon^{uV}_{\mu\mu}\) and (right) \(\varepsilon^{dV}_{ee}\) vs \(\varepsilon^{dV}_{\mu\mu}\).}
	\label{fig:NSI_eemumu_emu}
\end{figure*}

\begin{figure*}
	\centering
	\includegraphics[width=0.4\linewidth]{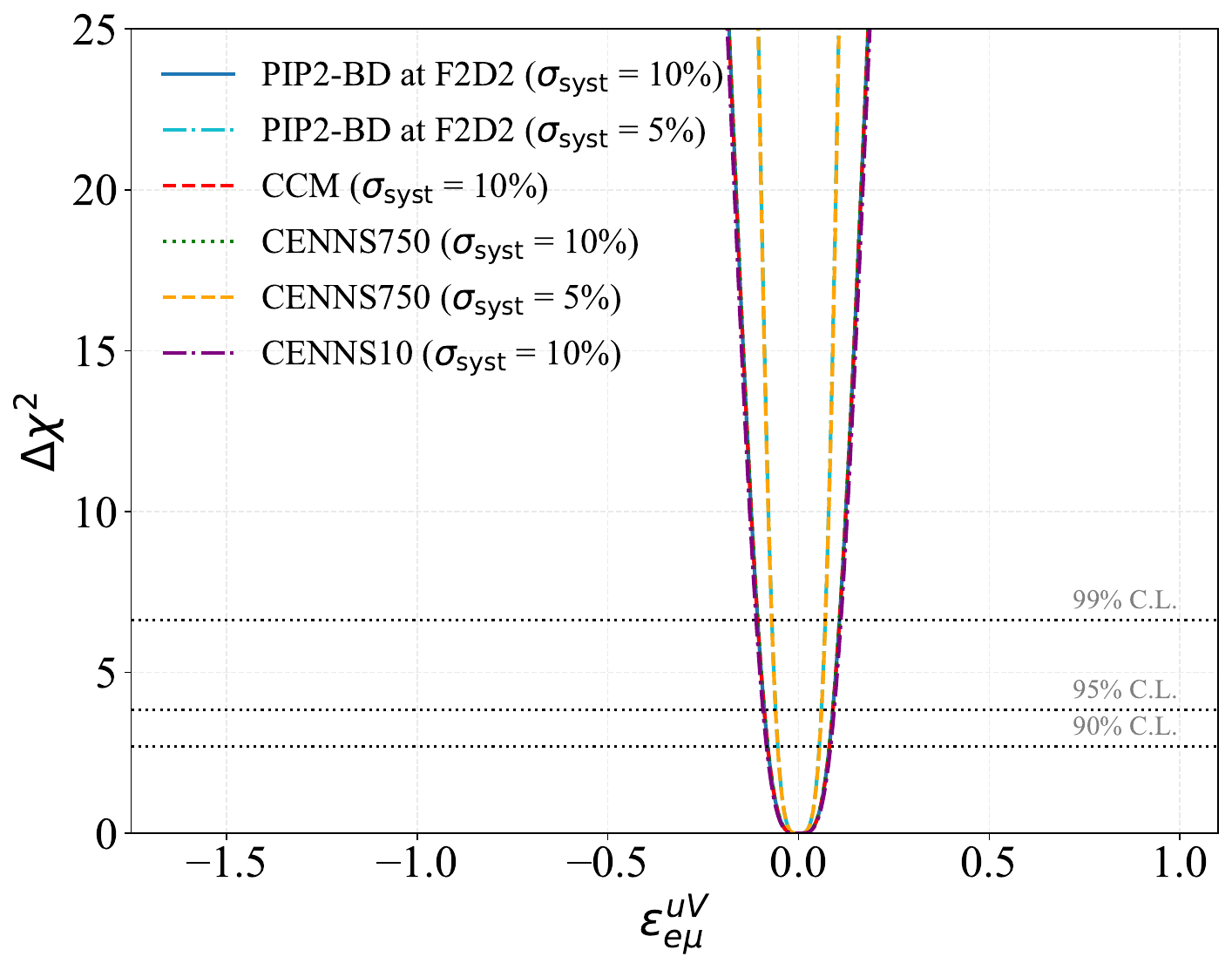}
    \includegraphics[width=0.4\linewidth]{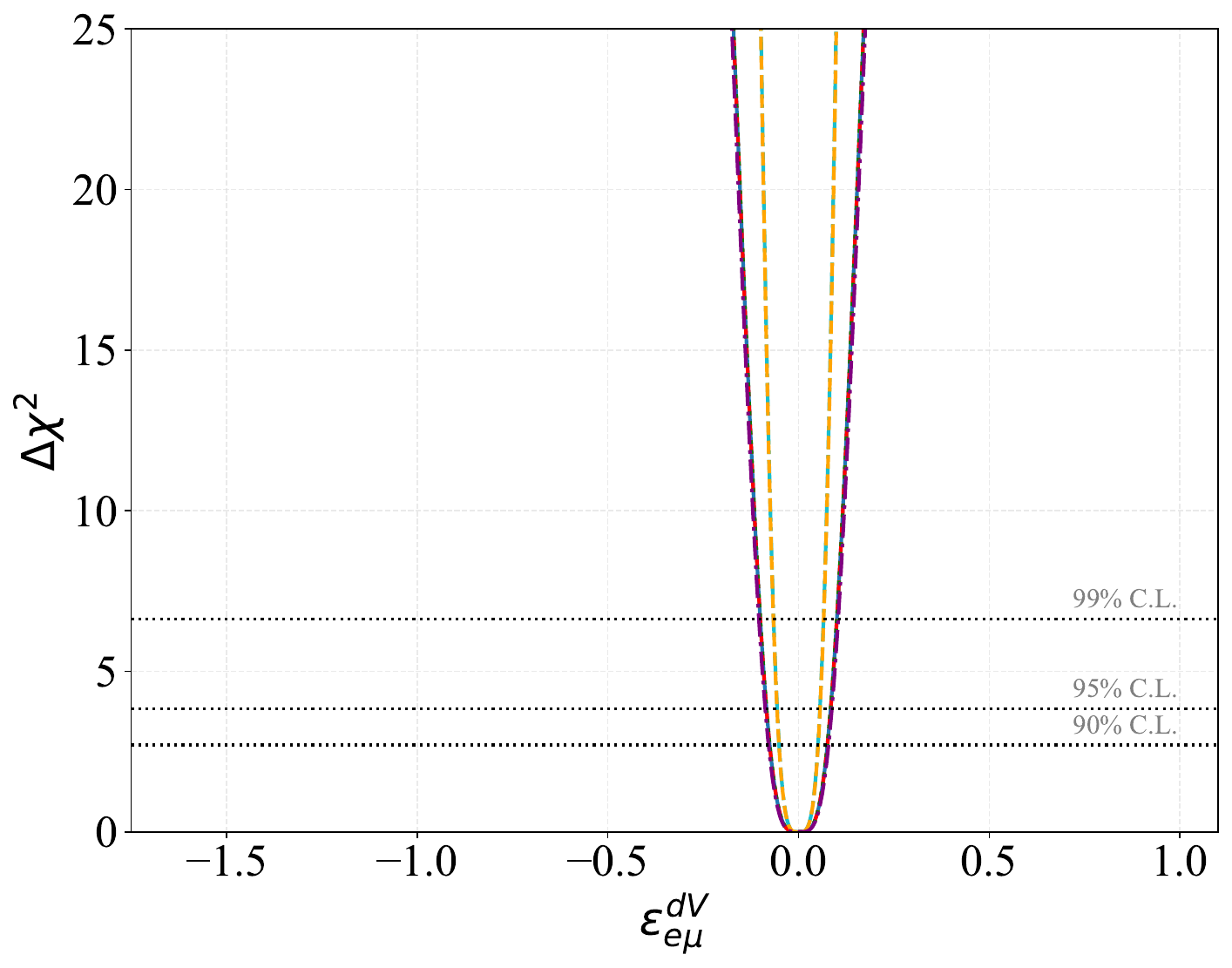}
	\caption{Sensitivity (\(\Delta \chi^2 = \chi^2-\chi^2_{\rm min}\)) on flavor off-diagonal NSI parameters: (left) \(\varepsilon^{uV}_{e\mu}\) and (right) \(\varepsilon^{dV}_{e\mu}\).}
	\label{fig:NSI_Uemu_Demu}
\end{figure*}

Here we study the sensitivity of CEvNS for different LAr experiments due to the following flavor diagonal NSI parameters: \(\varepsilon_{ee}^{uV},\,\varepsilon_{ee}^{dV},\,\varepsilon_{\mu\mu}^{uV},\text{and} \,\varepsilon_{\mu\mu}^{dV}\). The following flavor off-diagonal parameters: \(\varepsilon_{e\mu}^{uV},\text{and} \,\varepsilon_{e\mu}^{dV}\) are also studied here.  As in the previous analyses, we assume a three-year exposure for each configuration, and consider 5\% and 10\% systematic uncertainty scenarios. The sensitivity to the flavor-diagonal NSI parameters, considered one at a time, is shown in Appendix~\ref{app:sensitivity} (Fig.~\ref{fig:NSI_Uee_Dee_Umumu}). The resulting 90\% C.L. constraints are summarized in Tables~\ref{tab:NSI_ee_uV_dV} and~\ref{tab:NSI_mumu_uV_dV}. 
The allowed regions consist of two disjoint intervals, and the reported bounds correspond to the union of these ranges. The region allowed at 90\% CL for the flavor conserving NSI parameters: $\varepsilon_{ee}^{uV},\,\varepsilon_{ee}^{dV}$ is shown in Fig. \ref{fig:NSI_Umumu_vs_Dmumu}. On the same figure the allowed region for CHARM experiment from $\nu-N$ inelastic scattering is shown. Similar comparison for the flavor conserving NSI parameters: $\varepsilon_{\mu\mu}^{uV},\,\varepsilon_{\mu\mu}^{dV}$ is shown in Fig. \ref{fig:NSI_Umumu_vs_Dmumu}. The flavor diagonal comparison between $\varepsilon_{ee}^{uV},\,\varepsilon_{\mu\mu}^{uV}$ and $\varepsilon_{ee}^{dV},\,\varepsilon_{\mu\mu}^{dV}$ are shown in Fig. \ref{fig:NSI_eemumu_emu}. The sensitivity of the flavor off-diagonal NSI parameters one at a time are shown in Fig. \ref{fig:NSI_Uemu_Demu} and the best estimate for these parameters within 90\% CL is listed in Table \ref{tab:NSI_emu_combined}. The combined allowed region for these parameters: $\varepsilon_{e\mu}^{uV},\,\varepsilon_{e\mu}^{dV}$ is shown in Fig. \ref{fig:NSI_Umumu_vs_Dmumu}.
The sensitivity on the different parameters listed here for flux source CENNS10 is consistent with previous studies \cite{Liao:2017uzy, Papoulias:2017qdn, Giunti:2019xpr, Miranda:2020tif}. For the larger detectors such as the PIP2-BD, CENNS750 and CCM, the sensitivity on the parameters are improved in comparison to these previous analysis. The projected sensitivities demonstrate that these detectors can probe competitive NSI couplings benefiting from both their larger target mass and reduced systematic uncertainties. Such precision will allow detailed tests of flavor-conserving and flavor-diagonal NSI scenarios and, when combined other NSI probes, help resolve parameter degeneracies.

 \begin{table}
\centering
\begin{tabular}{@{}lcc@{}}
\toprule
\textbf{Experiments (Syst.\ Unc.)} & \boldmath$\epsilon_{e\mu}^{uV}$ & \boldmath$\epsilon_{e\mu}^{dV}$ \\
\hline
PIP2-BD at F2D2 (10\%)  & $[-0.080,\, 0.080]$ & $[-0.074,\, 0.074]$ \\
PIP2-BD at F2D2 (5\%)   & $[-0.054,\, 0.054]$ & $[-0.050,\, 0.050]$ \\
CENNS-10 (10\%)         & $[-0.084,\, 0.084]$ & $[-0.078,\, 0.078]$ \\
CENNS-750 (10\%)        & $[-0.080,\, 0.080]$ & $[-0.074,\, 0.074]$ \\
CENNS-750 (5\%)         & $[-0.054,\, 0.054]$ & $[-0.050,\, 0.050]$ \\
CCM (10\%)              & $[-0.080,\, 0.080]$ & $[-0.074,\, 0.074]$ \\
\hline
\end{tabular}
\caption{Estimated range for the off-diagonal NSI parameters \(\epsilon_{e\mu}^{uV}\) and \(\epsilon_{e\mu}^{dV}\) at 90\% C.L.\ for different argon-based
CEvNS experiments, assuming different levels of systematic uncertainty.}
\label{tab:NSI_emu_combined}
\end{table}


\section{Conclusions}\label{sec:conclusions}

Coherent elastic neutrino--nucleus scattering (CEvNS) has become a key experimental avenue for testing the SM and probing physics beyond it. In this work, we have investigated the prospects for argon-based CEvNS experiments to explore non-standard neutrino properties using stopped-pion neutrino sources at existing and future facilities. Our study focused on the CENNS-10 and CENNS-750 detectors at the Spallation Neutron Source at Oak Ridge Lab, the Coherent Captain Mills (CCM) detector at Los Alamos Lab, and the PIP2-BD detector at Fermilab's Facility for Dark Matter Discovery (F2D2).

Our analysis shows that argon detectors can achieve significant sensitivity to several low-energy electroweak parameters. For the weak mixing angle, we find that large-mass detectors such as CENNS-750 and especially PIP2-BD can achieve a precision competitive with or better than current low-energy determinations.  
The expected sensitivity would enable a meaningful test of the running of the weak mixing angle at low momentum transfers. In addition, we examined the capability of these experiments to probe the electromagnetic properties of neutrinos, including their magnetic moments and charge radii. The projected bounds improve upon or complement current limits from CEvNS and neutrino--electron scattering, highlighting the discovery potential of future high-statistics datasets in particular from CENNS-750 and PIP2-BD.

We also studied the reach of argon-based CEvNS detectors to constrain non-standard neutrino interactions (NSIs) with quarks. The expected sensitivities at PIP2-BD and CENNS-750 are competitive with, and in some cases surpass, those from existing experiments. They would provide critical input to global fits that combine CEvNS and other probes to help resolve parameter degeneracies. 

For the exposures considered here, particularly for PIP2-BD, several configurations already lie in a systematics-dominated regime, indicating that further improvements will primarily depend on reducing uncertainties associated with flux normalization, detector response, and nuclear modeling. Together, these results demonstrate that forthcoming argon-based CEvNS measurements will play a pivotal role in precision neutrino physics, offering a clean and coherent probe of weak interactions and an incisive window into new physics at the MeV scale.


\acknowledgments 

We thank Matt Toups and Jacob Zettlemoyer for discussions related to PIP2-BD at F2D2 at Fermilab. We thank
Gil Paz for reading the manuscript and giving helpful comments. S.C. would like to acknowledge the support of U.S. Department of Energy grant DE-SC0007983, the Visiting Scholars Award Program of the Universities Research Association, and support from the Fermilab Neutrino Physics Center Fellowship. This manuscript has been authored by Fermi Forward Discovery Group, LLC under Contract No. 89243024CSC000002 with the U.S. Department of Energy, Office of Science, Office of High Energy Physics.


\appendix

\section{Cross Section Comparison for Different Form Factors}
\label{app:ff_comparison}

In this appendix, we present additional results illustrating the impact of different nuclear form factor models on the CEvNS cross section. As discussed in the main text, the form factor encodes the finite spatial distribution of nucleons inside the nucleus and becomes increasingly important at higher momentum transfer.

\begin{figure}
    \centering
    \includegraphics[width=0.95\linewidth]{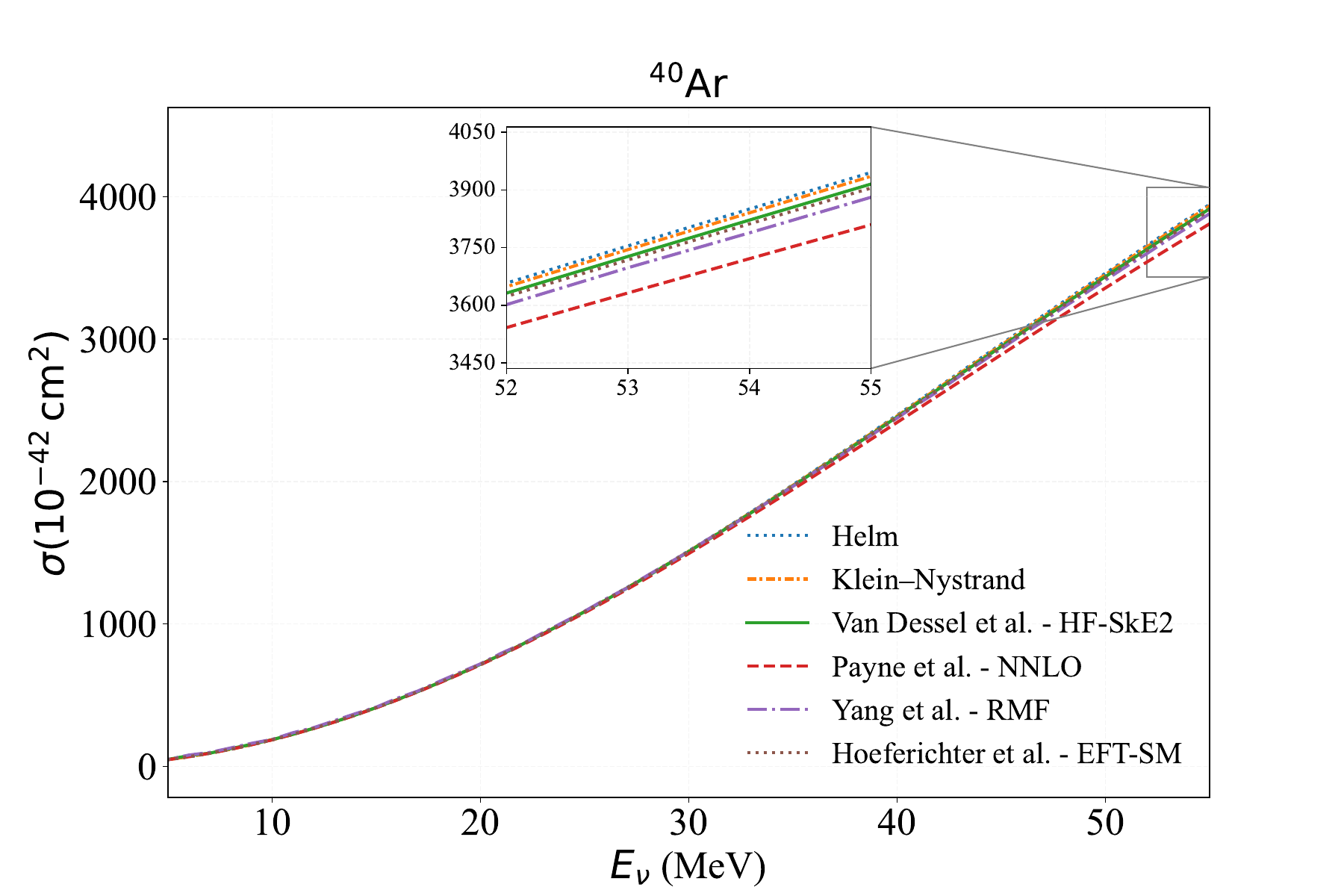}
    \caption{Total CEvNS cross section for $^{40}$Ar as a function of incident neutrino energy for different nuclear form factor models. The inset highlights the relative variation among the models at higher energies.}
    \label{fig:cs_cmprsn}
\end{figure}

The total CEvNS cross section for $^{40}$Ar, computed using several commonly used form factor parametrizations (outlined in Table~\ref{tab:events_formfactors}), is shown in Fig.~\ref{fig:cs_cmprsn}. At low neutrino energies, corresponding to small momentum transfer, all models yield nearly identical results, reflecting the coherent nature of the scattering process. In this regime, the form factor approaches unity and nuclear structure effects are negligible.

As the neutrino energy increases, deviations between the different form factor models become more pronounced. This behavior arises from the suppression of the cross section at larger momentum transfer, where the internal structure of the nucleus is resolved. The inset in Fig.~\ref{fig:cs_cmprsn} highlights these relative differences at higher energies.

Overall, while the choice of form factor has a negligible impact on the total rate at low energies, it introduces a modest model dependence at higher energies.

\section{Individual Sensitivity and Constraint Results}
\label{app:sensitivity}

For completeness, we present additional one-dimensional $\Delta\chi^2$ profiles for the observables discussed in the main text. These results provide a complementary view of the parameter sensitivities, highlighting features that are less apparent in the two-parameter analyses.

\begin{figure*}
	\centering
	\includegraphics[width=0.42\linewidth]{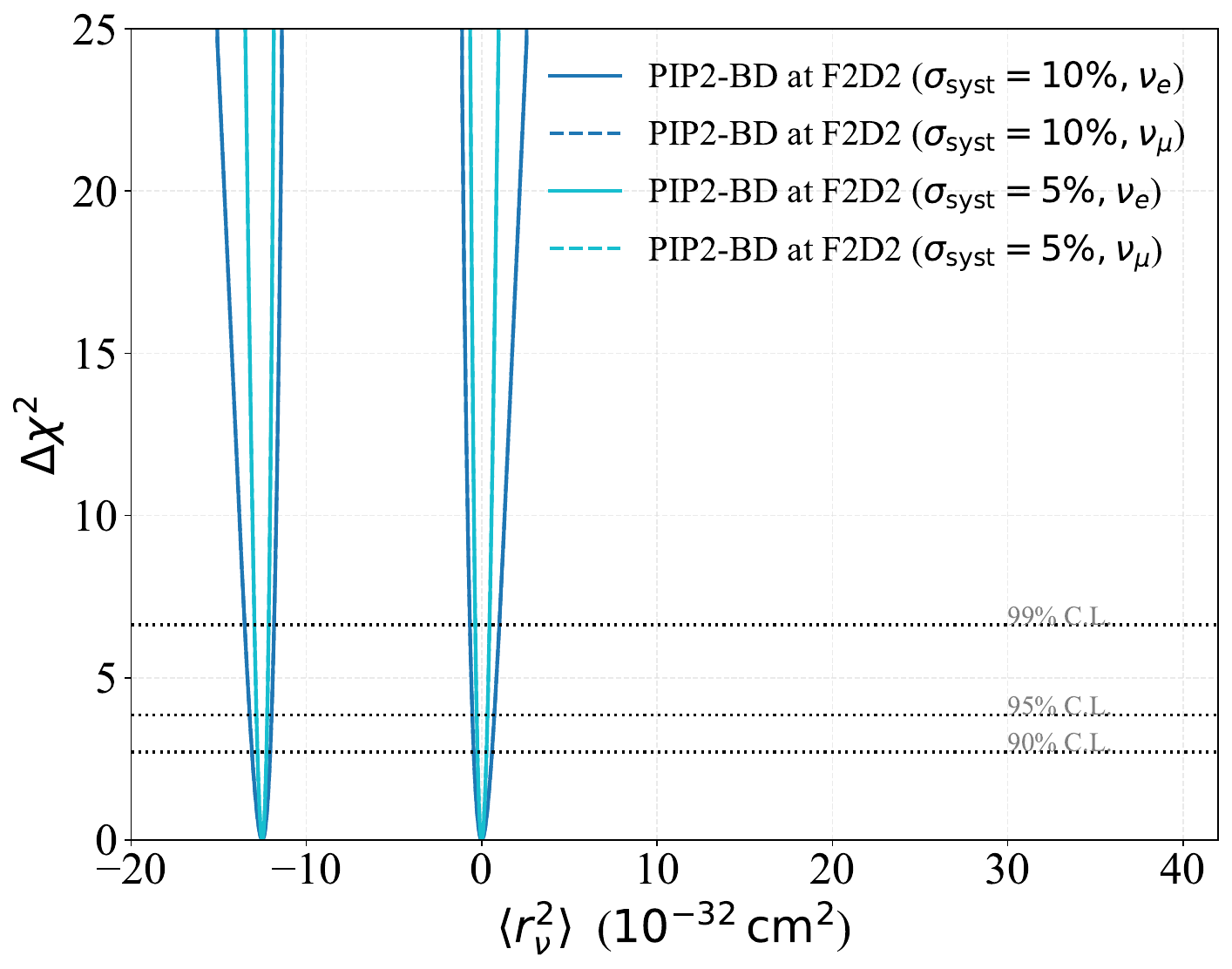}
    \includegraphics[width=0.42\linewidth]{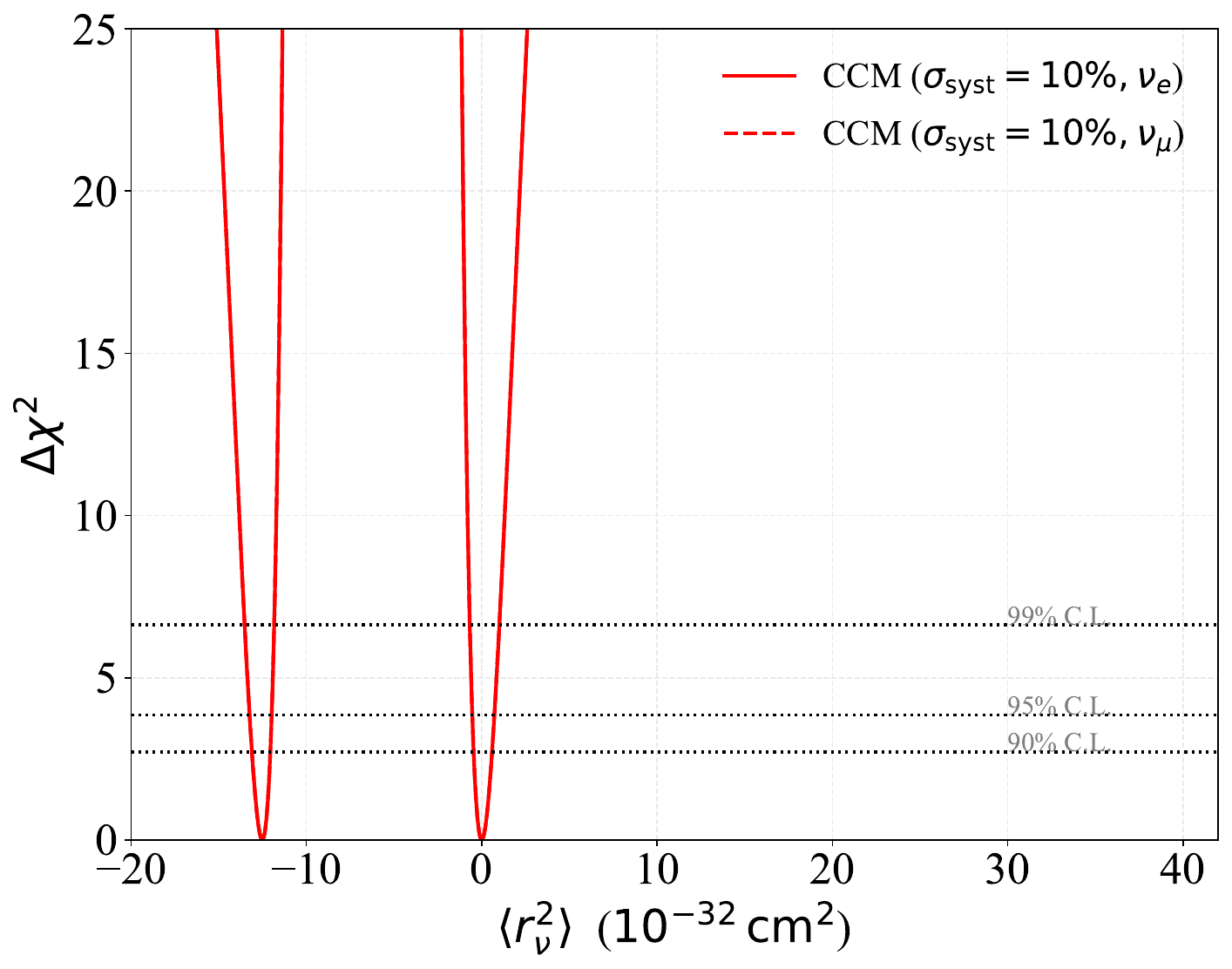}
    \includegraphics[width=0.42\linewidth]{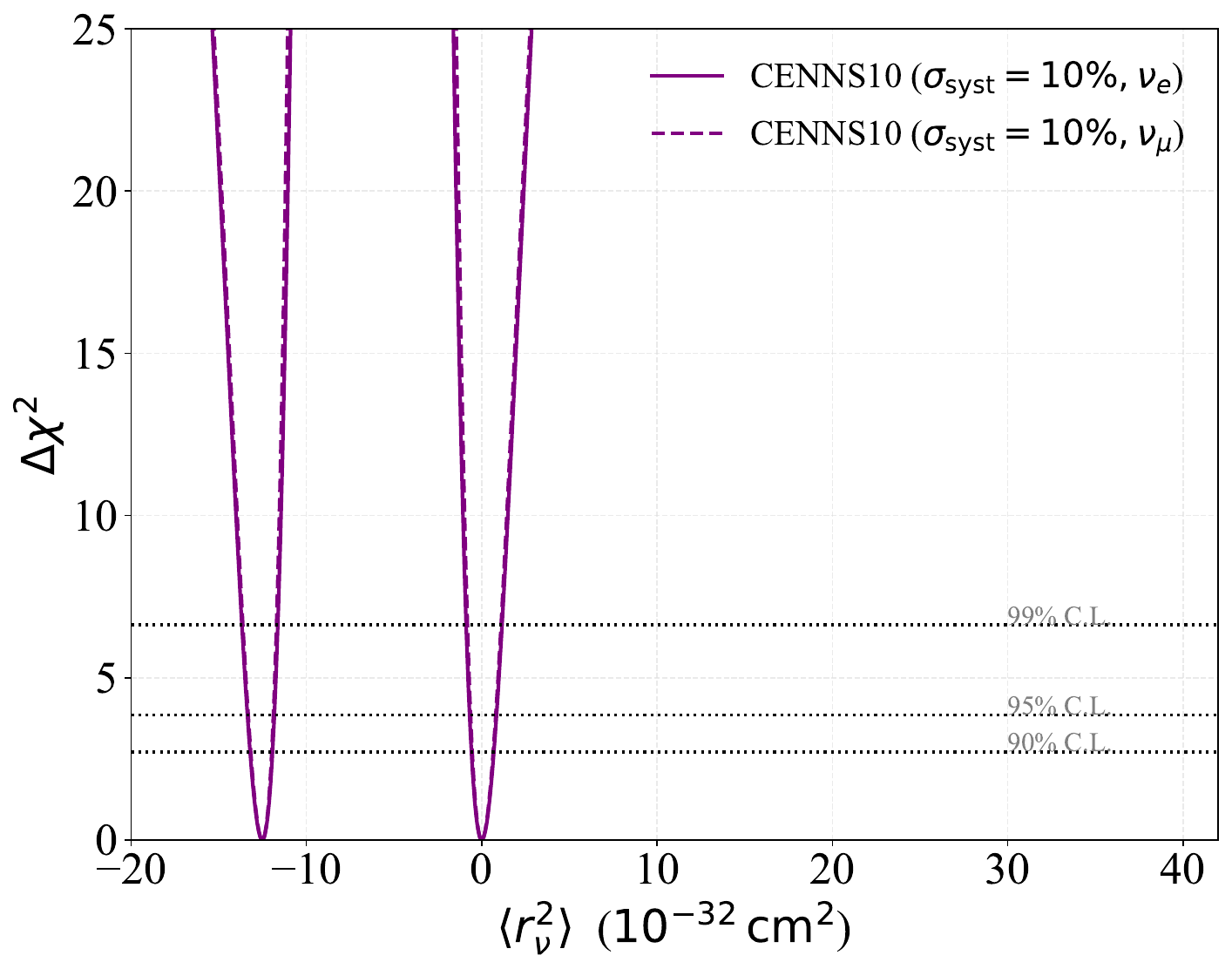}
    \includegraphics[width=0.42\linewidth]{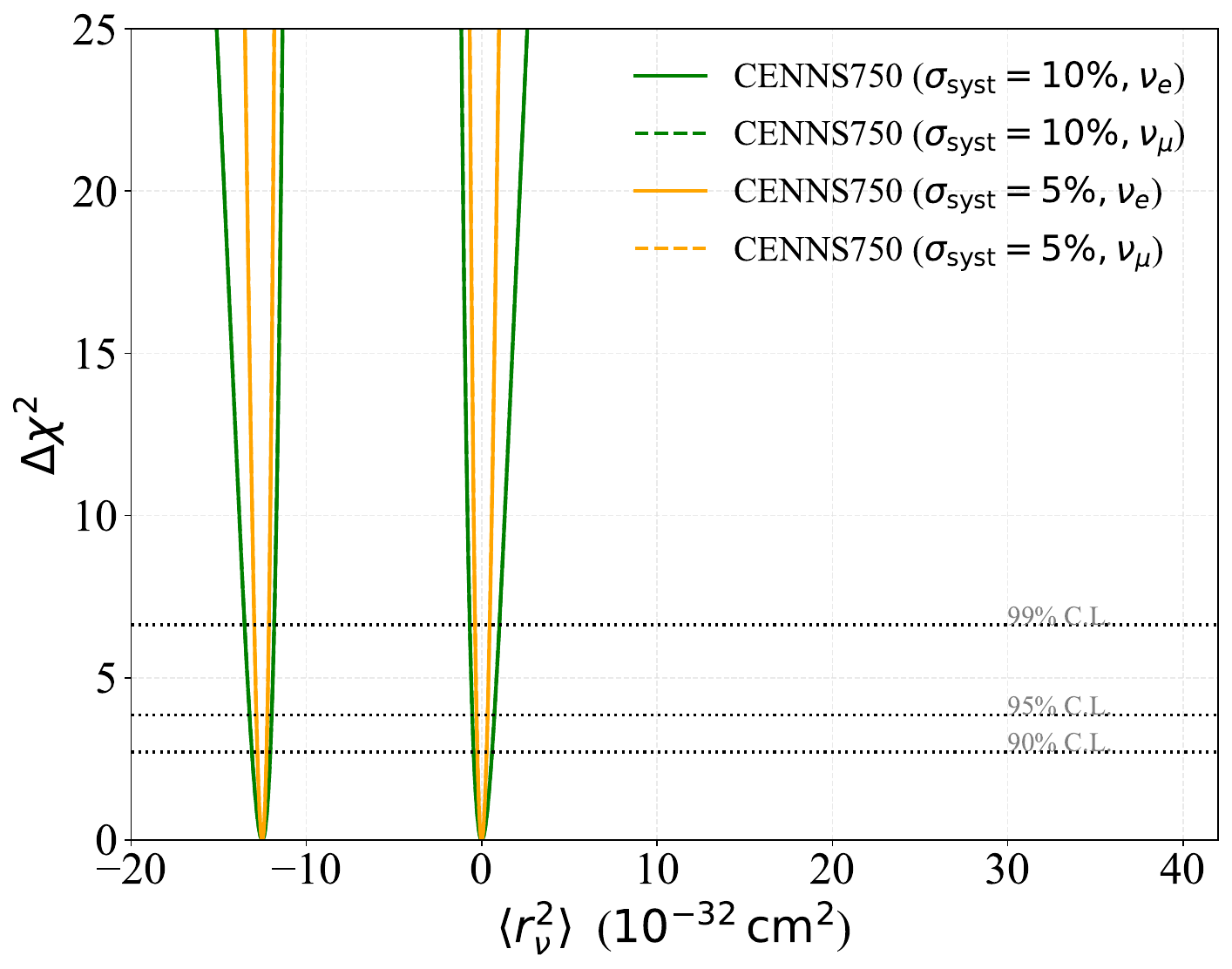}
	\caption{Sensitivity (\(\Delta \chi^2 = \chi^2-\chi^2_{\rm min}\)) on neutrino charge radius (\(\langle r^2 \rangle_{\nu_e},\,\langle r^2 \rangle_{\nu_\mu}\)) for the different flux sources: PIP2-BD at F2D2 with \(\sigma_{\rm{syst}}\) = 10\% and 5\% (top left), CCM with \(\sigma_{\rm{syst}}\) = 10\% (top right),  CENNS10 with \(\sigma_{\rm{syst}}\) = 10\% (bottom left), and CENNS750 with \(\sigma_{\rm{syst}}\) = 10\% and 5\% (bottom right).}
	\label{fig:Charge_radius_Flvr_all_Sources}
\end{figure*}

\begin{figure}
    \centering
    \includegraphics[width=0.8\linewidth]{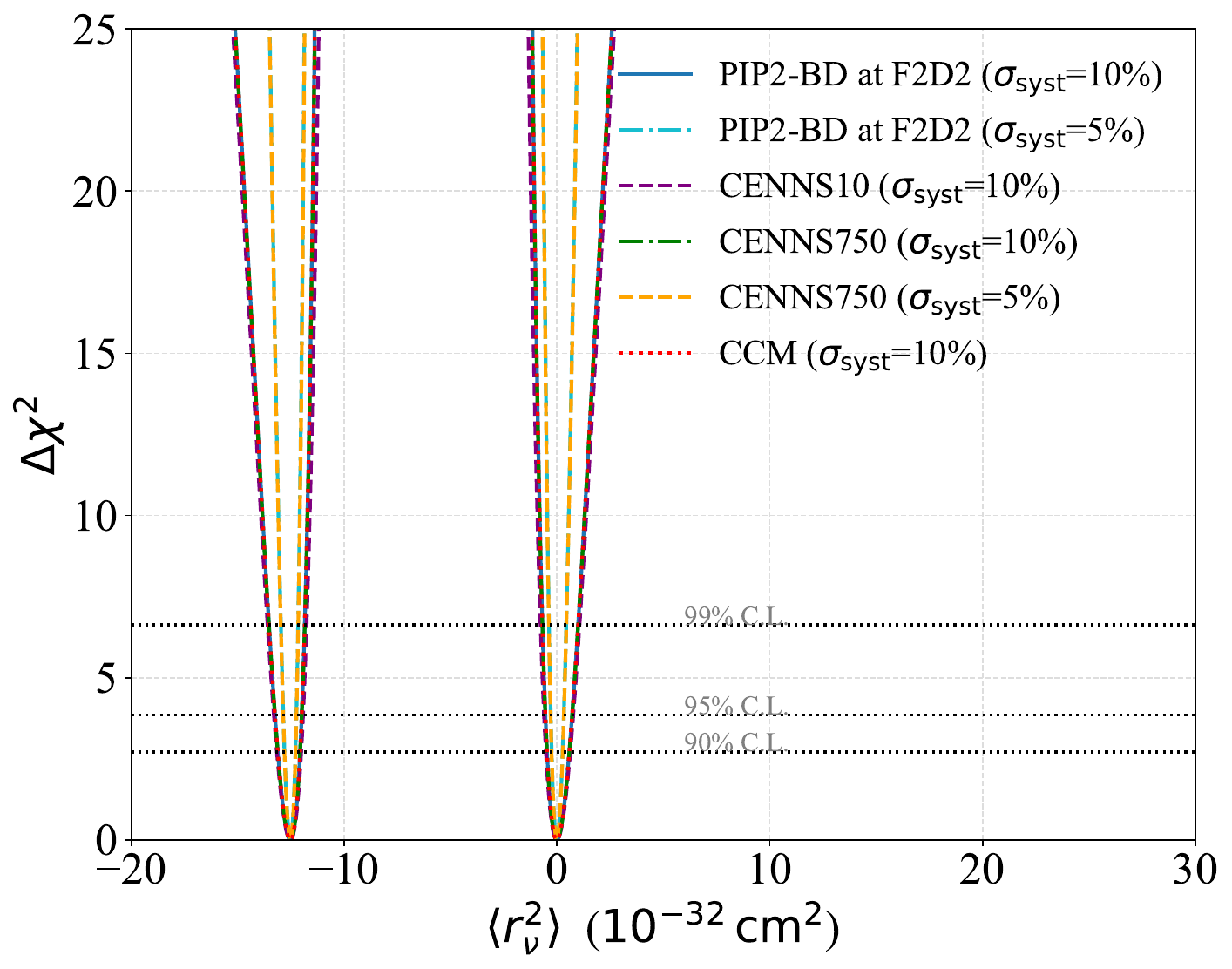}
    \caption{Sensitivity (\(\Delta \chi^2 = \chi^2-\chi^2_{\rm min}\)) to the universal effective neutrino charge radius for the different flux sources.}
    \label{fig:Charge_Radius}
\end{figure}

\begin{figure*}
	\centering
	\includegraphics[width=0.42\linewidth]{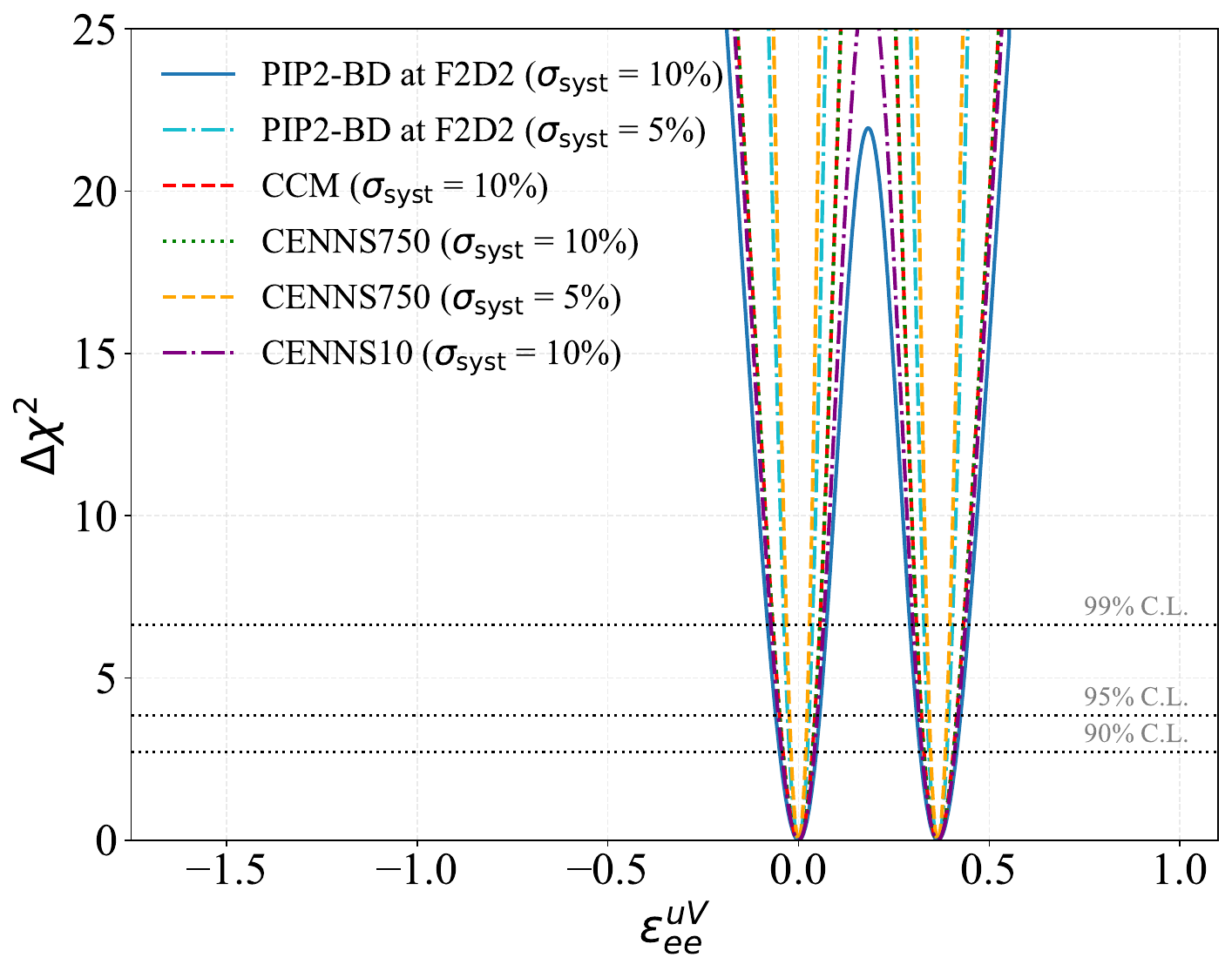}
    \includegraphics[width=0.42\linewidth]{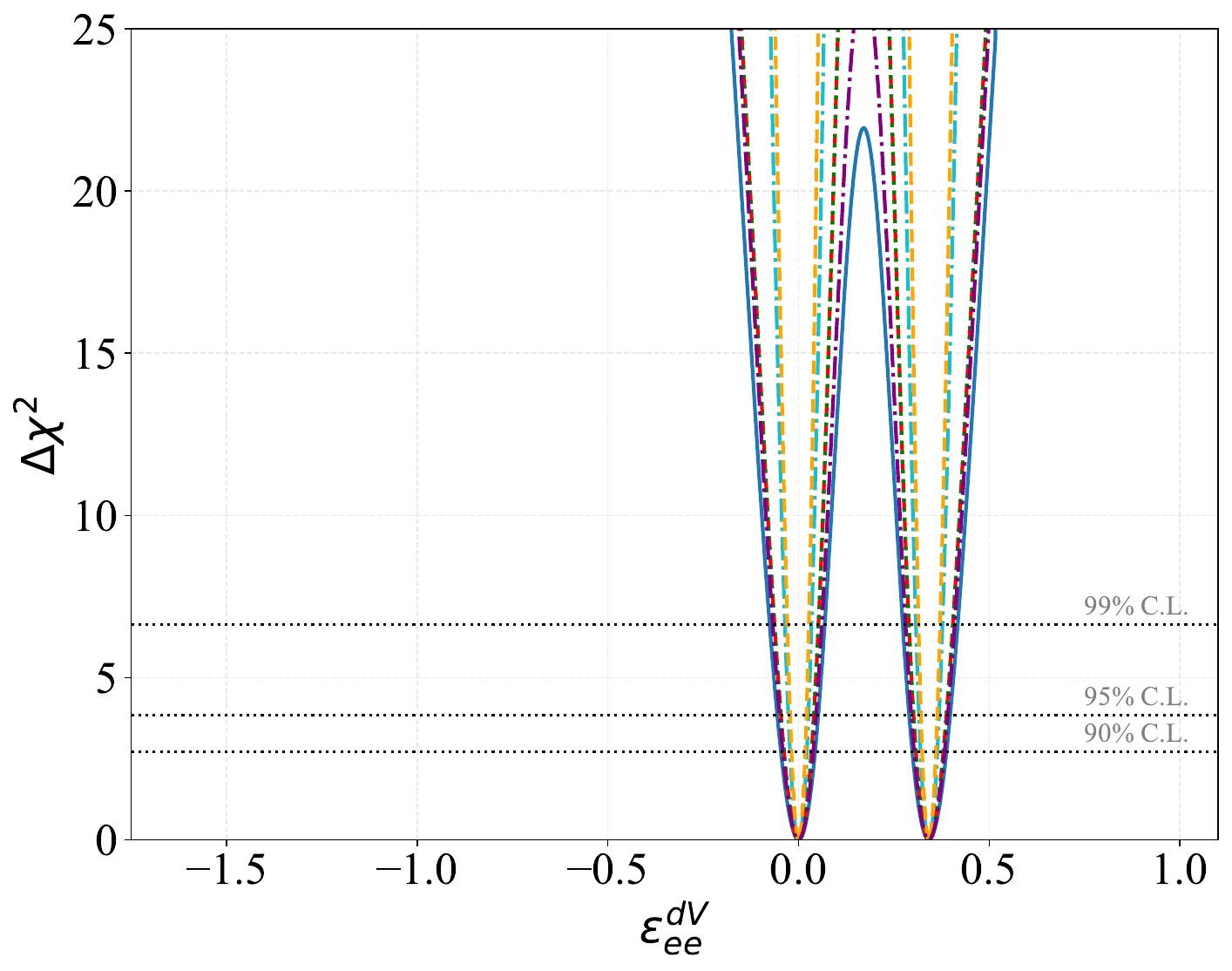}
    \includegraphics[width=0.42\linewidth]{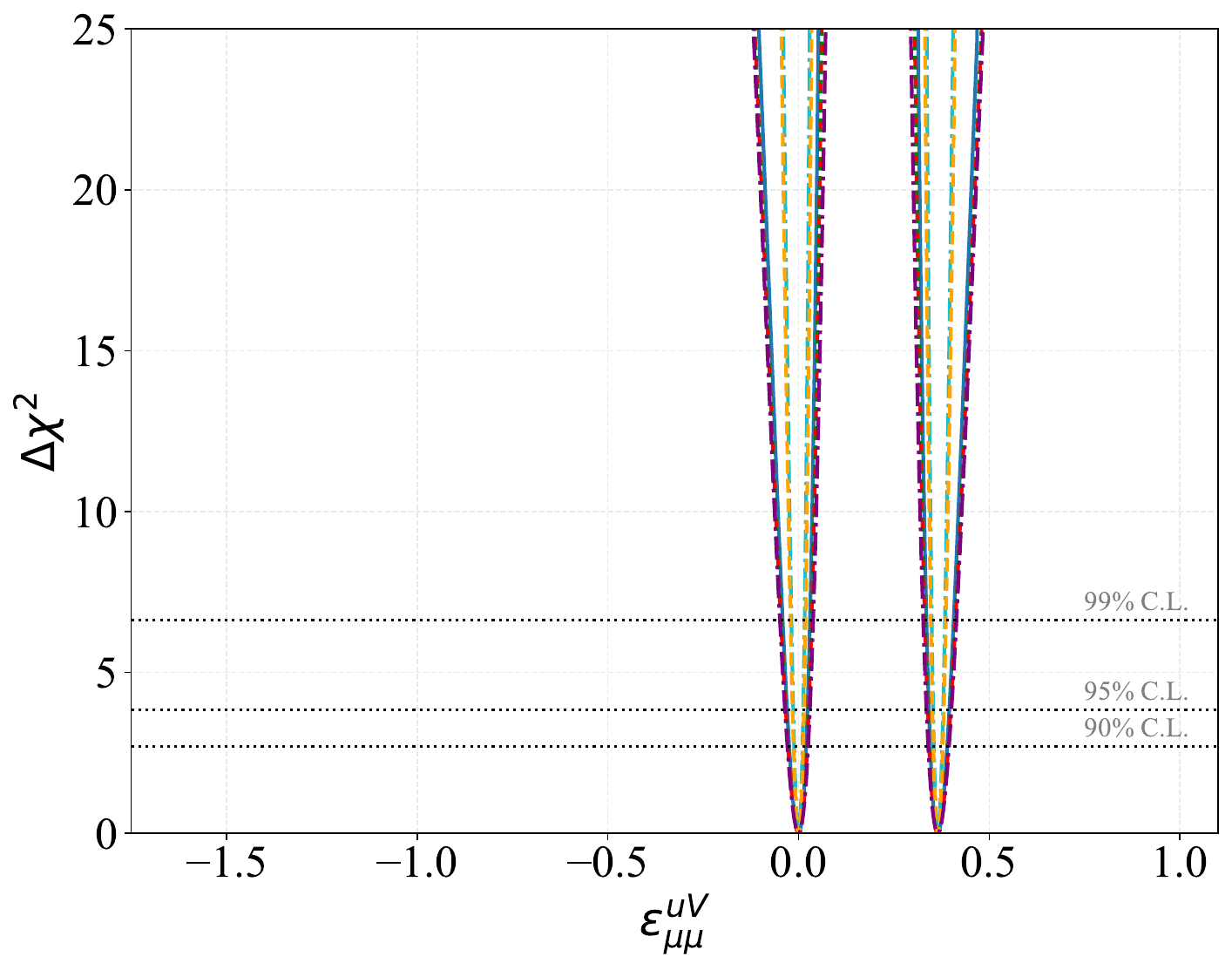}
    \includegraphics[width=0.42\linewidth]{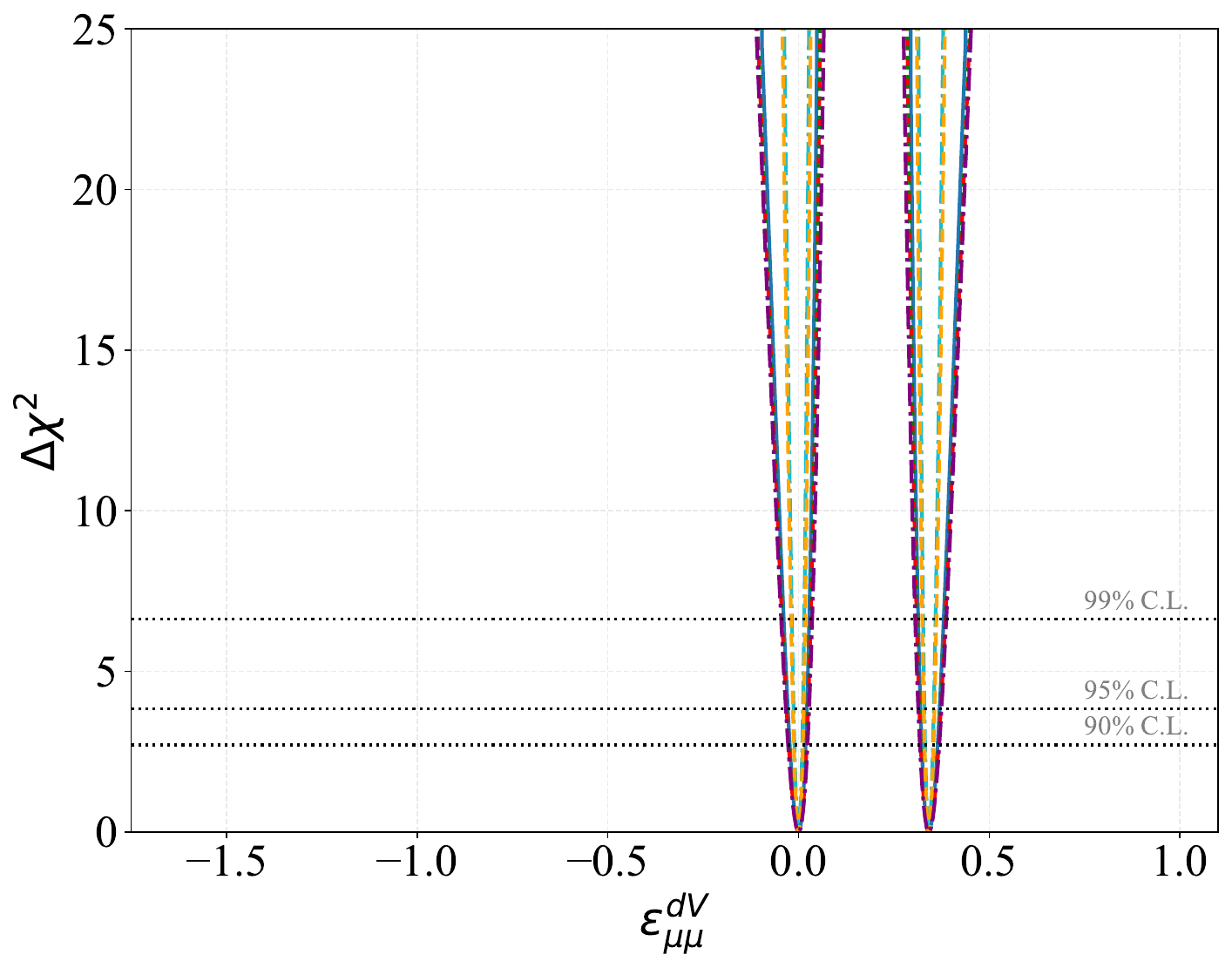}
	\caption{Sensitivity (\(\Delta \chi^2 = \chi^2-\chi^2_{\rm min}\)) on flavor diagonal NSI parameters: \(\varepsilon^{uV}_{ee}\) (top left), \(\varepsilon^{dV}_{ee}\) (top right), \(\varepsilon^{uV}_{\mu\mu}\) (bottom left), and \(\varepsilon^{dV}_{\mu\mu}\) (bottom right).}
	\label{fig:NSI_Uee_Dee_Umumu}
\end{figure*}

The sensitivity to the flavor-dependent neutrino charge radius is shown in Fig.~\ref{fig:Charge_radius_Flvr_all_Sources}, while the sensitivity to the universal (flavor-independent) effective charge radius is shown in Fig.~\ref{fig:Charge_Radius}. The corresponding constraints on flavor-diagonal NSI parameters are shown in Fig.~\ref{fig:NSI_Uee_Dee_Umumu}. In both cases, the $\Delta\chi^2$ profiles exhibit the characteristic two-fold degeneracy arising from the quadratic dependence of the CEvNS rate on the effective couplings, leading to disjoint allowed regions at a given confidence level.


\bibliography{references}


\end{document}